\documentclass{aa}  

\usepackage{booktabs}
\usepackage[table,xcdraw]{xcolor}
\usepackage{graphicx}
\usepackage{amsmath}
\usepackage{txfonts}
\usepackage{lipsum}
\usepackage{subcaption}   
\usepackage{lscape}             
\usepackage{placeins}           
\newcommand{\bs}[1]{\boldsymbol{#1}}
\newcommand{\hbs}[1]{\hat{\boldsymbol{#1}}}
\newcommand{\vbs}[1]{\bar{\boldsymbol{#1}}}

\usepackage{hyperref}

\begin{document}

   \title{Triple-induced mergers of black hole binaries}
   \subtitle{A comprehensive look at the role of stellar evolution, dynamical stability, and spin evolution.}

   \author{C. W. Bruenech\inst{1}
        \and S. Toonen\inst{1}
        \and T. Boekholt\inst{1}
        \and A. Dorozsmai\inst{2}
        }

   \institute{Anton Pannekoek Institute for Astronomy\\
             Science Park 904, 1098 XH Amsterdam, the Netherlands\\
             \email{c.w.bruenech@uva.nl}
            \and
            {National Astronomical Observatory of Japan}\\
                National Institutes of Natural Sciences, 2-21-1 Osawa, Mitaka, Tokyo 181-8588, Japan}

   \date{Received September 30, 20XX}

\abstract
{Through observations of gravitational waves, mergers of black holes (BHs) have been shown to be a ubiquitous phenomenon in the Universe. However, uncertainties remain in our theoretical understanding of the evolution of the BHs prior to their merger. Black hole progenitors are seemingly born primarily in triples or higher-order multiples, and the presence of a tertiary object can perturb a black hole binary (BHB) enough to precipitate a merger.}
{We aim to provide a detailed overview of BH mergers in wide triples by evolving systems from stellar birth to BH merger, using non-orbit-averaged methods for both the dynamics and BH spins from the point of BHB formation, and including triples that become dynamically unstable due to stellar evolution. This allows us to obtain a more complete picture of BH mergers in triples, shedding light on the initial orbital properties required to produce a merger in a triple, as well as the final properties of the merging BHBs.}
{We used population synthesis coupled with orbit-averaged descriptions of triple dynamics to evolve a population of triples with initially wide orbits and massive progenitors. Wide orbits were chosen to avoid stellar interaction prior to BH formation. Systems that remain bound and form an inner BHB were evolved using a direct n-body code with post-Newtonian terms up to an order of 2.5. We also simulated the precession of the BH spin vectors by coupling the n-body solver with the differential equations for the spins.}
{Forming a wide, bound BHB with a tertiary companion requires the system to survive pre-BH stellar evolution and at least two supernovae in the inner binary. Consequently, only a small fraction of the massive triple population are born with the parameters required to achieve this configuration. For the dynamically stable triples with inner BHBs, only the systems with inclinations close to $90^\circ$ can excite the eccentricity in the inner binary to values high enough for gravitational waves (GWs) to dissipate sufficient energy to merge the system with an estimated merger rate density of $\sim 5$ Gpc$^{-3}$ yr$^{-1}$. Mergers also occur in triples that become dynamically unstable at a rate of $\sim 1.4$ Gpc$^{-3}$ yr$^{-1}$. At the point of entering the 10 Hz gravitational wave frequency band, the merging inner binaries exhibit eccentricities between $10^{-4}$ and $10^{-2}$. Finally, we find that the final effective spin of a BHB that merges through this channel can display a wide range of values between $-1$ and $1$, with a slight tendency towards $\chi_\text{eff} \approx 0$. This is a result of the strong three-body dynamics experienced by the merging triples before the inner binary begins to shrink due to GW emission. The inner angular momentum can explore the full phase space before the binary rapidly shrinks and decouples from the tertiary, effectively freezing out the effective spin to its value at the time of the highest inner eccentricity.}
{}


   \maketitle
   \nolinenumbers

\section{Introduction}

With the recent release of version 4.0 of the gravitational-wave transient catalogue (GWTC) by the LIGO collaboration, the total number of observed compact binary mergers sits at 218, with the majority of these consisting of binary black hole mergers \citep{abac_gwtc-40_2025}. Forming a black hole binary (BHB) with a separation small enough to merge within the age of the Universe has shown to be a highly non-trivial problem and has generally been divided into two main formation channels: isolated binary evolution and dynamical interactions. The former is often split into classical binary evolution (e.g. \citet{bethe_evolution_1998, dominik_double_2012, marchantEvolutionBinaryStars2025}) and chemically homogenous evolution \citep{mandel_new_2016, de_mink_chemically_2016, marchant_new_2016, du_buisson_cosmic_2020, sharpe_investigating_2024, dorozsmai_stellar_2024}. Meanwhile, the dynamical formation channel includes scenarios such as mergers in dense stellar environments including globular clusters \citep{wen_eccentricity_2003, antonini_black_2014, rodriguez_post-newtonian_2018, fragione_black_2018} and mergers in active galactic nuclei (AGNs) through binary-single interactions \citep{yang_investigating_2019, samsing_agn_2022, 2023MNRAS.524.2770R, grishin_effect_2024, 2024MNRAS.531.4656W, gilbaum_how_2025}, gas interactions \citep{lai_circumbinary_2023, boekholt_jacobi_2023, mckernan_constraining_2024}, and BHBs in hierarchical triples \citep{antonini_black_2014, silsbee_lidov-kozai_2017, liu_spin-orbit_2017, rodriguez_triple_2018, antonini_precessional_2018, liu_black_2018, rodriguez_triple_2018, yu_spin_2020, martinez_mass_2022}, the latter of which can be formed both dynamically in clusters and in the field. In this study we focused on hierarchical field triples. We labelled BHB mergers in these systems as dynamical since the merger occurs exclusively as a result of triple dynamics.

Massive main-sequence stars, which are the progenitors to black holes (BHs), have been shown to predominantly reside in triples or higher-order multiples \citep{sana_binary_2012, moe_mind_2017, offner_origin_2023}, which begs the question of whether or not high-order gravitational dynamics may play a role in the evolution of the stellar constituents and subsequent BH remnants. In hierarchical triples, where an inner binary is orbited by a distant tertiary body, the inner binary may be driven to extremely high eccentricities due to perturbations from the outer companion (see Sect. \ref{sec:primer}). This effect, coupled with energy dissipation through gravitational waves (GWs) could induce mergers in BHBs that would otherwise be too wide to merge within the age of the Universe. This dynamical, triple-induced BHB merger channel has been shown, through various theoretical studies, to produce mergers in rates of $\sim0.2 - 25$ Gpc$^{-3}$ yr$^{-1}$ \citep{antonini_black_2014, silsbee_lidov-kozai_2017, rodriguez_triple_2018, martinez_mass_2022}. As a consequence of the strong perturbations by the tertiary on the inner BHB, the merging binaries are more eccentric in the later stages of their inspiral compared to similar systems from isolated binary evolution, with reported eccentricities as high as $0.999$ \citep{silsbee_lidov-kozai_2017} when they enter the 10 Hz GW frequency band. For spinning black holes, triple-induced mergers can also cause the black hole spin vectors to misalign with respect to the binary angular momentum \citep{liu_spin-orbit_2017, antonini_precessional_2018, liu_black_2018, rodriguez_triple_2018, yu_spin_2020, fragione_effective_2020}, resulting in a broad distributions of possible effective spin values during the final inspiral. High eccentricity and spin-orbit misalignment could therefore be observable properties that may indicate that a system merged via the perturbations of a tertiary \citep{stegmann_distinguishing_2025}. 

In addition, there is a related channel involving triples that lose their hierarchical structure -- due to, for example, mass loss from winds or explosive events -- at which point the system becomes dynamically unstable, enabling rich gravitational dynamics. During this time, the triple constituents can experience close encounters wherein GWs may remove enough energy to precipitate a compact object merger. Dynamically unstable triples have previously been excluded from studies on triple-induced BH mergers as they cannot be simulated using orbit-averaged techniques.

Compared to isolated binary evolution, mergers of BHBs in triples can produce systems with higher eccentricities at the point of the entering the observable sensitivity band of current GW observatories. Additionally, perturbations from the tertiary can lead to spin-orbit misalignment at the time of merger, which can result in a substantial recoil kick \citep{campanelli_large_2007} and non-zero effective spins. These observables can be beneficial in not only increasing our understanding of massive stars but also shed more light on which evolutionary channel is dominant in forming BHB mergers. Analysis of GWTC have shown that the majority of merging BHBs have spin vectors aligned with the binary orbit, implying that they likely form due to isolated binary evolution \citet{abac_gwtc-40_2025}. However, a substantial fraction still exhibits signs of misaligned and/or anti-aligned spins, which could imply that they are the result of dynamical effects such as three-body dynamics. An analysis of BHB mergers from the first three LVK observing runs found evidence of eccentric orbits in two out of 90 events \citep{planas_reanalysis_2025}, supporting the theory that a significant fraction of BHB mergers may be formed in dynamical environments. Similarly, the BH-neutron star merger event GW200105 has also recently been shown to potentially have an eccentricity of $0.12$ at the 20 Hz frequency band \citep{planas_first_2025}, suggesting that the merger occurred as a result of dynamical interactions.

In this work we present a detailed study on the formation of BHB mergers in both hierarchical triples and in triples that become dynamically unstable due to internal stellar evolution. By following the evolution of massive star systems from stellar birth to merger using a combination of secular and detailed simulations, we acquired a complete overview of the evolution of BHBs that merged due to triple dynamics. More specifically, we used a population synthesis approach to evolve the stellar progenitors from the zero age main sequence (ZAMS) to the formation of an inner BHB before switching to a direct n-body code with post-Newtonian terms up to an order of $2.5$. We also evolve the black hole spins using direct (non-orbit-averaged) spin precession equations to obtain an accurate overview of the evolution of the spin vectors from the formation of the BHB. In particular, we examine whether triples that become dynamically unstable can produce BHB mergers, an aspect omitted from previous studies. In Sect. (\ref{sec:primer}) we give a short introduction to hierarchical triples and the mechanism behind the eccentricity excitations, in addition to a brief overview on dynamically unstable triples. To follow, we present our methodology in Sect. (\ref{sec:method}, where we mainly detail the different codes used to model the various parts of the evolution. In Sect. (\ref{sec:results}) we provide an overview of our simulation results, while in Sect. (\ref{sec:discussion}) we compare our results to other studies of the same channel, including short discussions on the uncertainties of our model assumptions and the codes utilized.

\section{A primer on hierarchical triples}\label{sec:primer}

Hierarchical triples are stellar systems consisting of three bodies arranged as two nested binaries: an inner binary whose centre of mass (COM) is in a binary configuration with a tertiary object. The term 'hierarchical' arises because, when the separation between the tertiary and the inner COM is much larger than the separation between the bodies in the inner binary, the system can be approximated as a hierarchy of two nested Kepler orbits. We can consequently describe the system using a set of orbital elements. The most important of these include the semi-major axes $a_\text{in}$ and $a_\text{out}$, eccentricities $e_\text{in}$ and $e_\text{out}$, and the mutual inclination $i_\text{mut}$, defined as the angle between the inner and outer orbital angular momentum vectors $\vec{L}_\text{in}$ and $\vec{L}_\text{out}$. A hierarchical triple with a non-zero mutual inclination experiences a torque on the inner binary from the tertiary, allowing the two orbits to exchange angular momentum, which can excite eccentricity while decreasing inclination. This process, known as the von Zeipel-Lidov-Kozai (ZKL) mechanism \citep{von_zeipel_sur_1910, lidov_evolution_1962, kozai_secular_1962}, occurs in periodic cycles on timescales determined mainly by the ratio $P_\text{out}^2/P_\text{in}$ (see \citet{kinoshita_analytical_1999} for the full expression), with larger ratios resulting in longer timescales. The maximum eccentricity achieved during a ZKL cycle is determined mainly by the inclination, with values close to $90^\circ$ producing the highest values of $e_\text{in}$.
When the semi-major axis ratio $a_\text{out}/a_\text{in}$ is large, the outer orbit is close to circular ($e_\text{out} \approx 0$), and the inner binary masses are near equal. Hence a hierarchical triple with a non-zero inclination is said to experience classic or regular ZKL cycles wherein the peak eccentricity $e_\text{max}$ during each ZKL cycle remains constant. However, if any combination of these criteria is unmet, the triple can experience richer dynamics. These may include a change in the maximum eccentricity (typically on timescales longer than classical ZKL timescales), orbital flips (where the inner binary transitions between retrograde and prograde, or vice versa), and eccentricities close to unity. In such cases, the triple is said to experience eccentric ZKL cycles (see \citet{naoz_eccentric_2016} for an overview of this mechanism). We can quantify the strength of this effect with the octupole parameter $\epsilon_\text{oct}$, defined as

\begin{equation}\label{eq:eps_oct}
    \epsilon_\text{oct} = \frac{a_\text{in}}{a_\text{out}}\frac{m_1 - m_2}{m_1 + m_2}\frac{e_\text{out}}{1 - e_\text{out}^2}.
\end{equation}

While there is no hard boundary between ZKL and eccentric ZKL, a triple system is typically assumed to undergo eccentric ZKL when $\epsilon_\text{oct} \gtrsim 10^{-2}$.

If the gravitational influence of the tertiary on the inner binary becomes too strong -- due to either a decreasing semi-major axis ratio $a_\text{out}/a_\text{in}$, an increasing outer eccentricity $e_\text{out}$, or both -- the hierarchy effectively breaks down. Consequently, the system can no longer be accurately described as nested Kepler orbits with the corresponding orbital elements. Changes in the orbital elements can result from stellar evolution, such as mass loss in the inner binary due to winds or supernova explosions. Several criteria exists for determining the stability of a triple system \citep{vynatheyaAlgebraicMachineLearning2022a, hayashiDynamicalDisruptionTimescales2022, tory_empirical_2022}, and in this work we used the criterion from \citet{mardling_tidal_2001}:

\begin{equation}\label{eq:mardling_aarseth}
    \frac{a_\text{out}}{a_\text{in}}\bigg|_\text{crit} = \frac{2.8}{1 - e_\text{out}}\left(1 - \frac{0.3i_\text{mut}}{\pi}\right) \left(\frac{(1 + q_\text{out})(1 + e_\text{out})}{\sqrt{1 - e_\text{out}}} \right)^{2/5}.
\end{equation}

A triple is classified as unstable if its semi-major-axis ratio $a_\text{out}/a_\text{in}$ becomes smaller than the critical value in Eq. (\ref{eq:mardling_aarseth}).

Earlier work on dynamically unstable triples using n-body simulations has found that the vast majority of these systems quickly disintegrate, either ejecting one of the components or resulting in a collision between two bodies \citep{hamers_return_2022, toonen_stellar_2022, bruenech_massive_2025}. Additionally, a small fraction of systems remain hierarchical over long timescales while lingering near the boundary of stability, exhibiting strong ZKL-like dynamics, including orbital flips and extreme eccentricity excitations. Unstable triples have therefore been shown to induce extremely close passages between bodies, which, when coupled with GR effects, could potentially lead to GW captures and BH mergers.

\section{Overview of numerical methods}\label{sec:method}

Our method for studying mergers of binary black holes with tertiary companions can be summarized as follows. We first simulated a population of triples with massive stars using the triple population synthesis code \texttt{TRES}. The subset of systems that form a bound inner BHB were then transferred to the direct n-body code \texttt{Syzygy} \citep{bruenech_massive_2025}, which further simulated the systems with the inclusion of post-Newtonian terms until a stopping condition was triggered. We simulated both dynamically stable triples and 
triples that become dynamically unstable due to stellar evolution. The triple systems that result in a merger were then simulated once again, this time with the inclusion of spin evolution. Finally, to reduce the total computational cost of the simulations, we simulated the final orbital evolution of the merging systems using the orbit-averaged GR equations from \citet{peters_gravitational_1964} once the inner binary dynamically decoupled from the tertiary. What follows is a more detailed overview of each step of the process.

\subsection{Generating wide binary black holes with tertiary companions using TRES}\label{sec:tres_method}

\texttt{TRES} \citep{toonen_evolution_2016} is a triple population synthesis code that combines stellar evolution with orbit-averaged triple dynamics to rapidly simulate the evolution of hierarchical triples from the ZAMS. TRES operates by coupling different codes for evolving different aspects of a triple system. The orbits in a given triple system were evolved in time by solving a set of ordinary differential equations (ODEs) that describe the time evolution of the orbital elements $a_\text{in}$, $a_\text{out}$, $e_\text{in}$, $e_\text{out}$, $\omega_\text{in}$, $\omega_\text{out}$, $h_\text{in}$, and $\theta \equiv cos(i_\text{mut})$, where $\omega$ denotes the argument of pericentre and $h$ denotes the angular momentum. Each ODE is a sum of effects that contribute to the change of a given parameter. For example, the total time derivative of the inner eccentricity, $\dot{e}_\text{in}$, includes changes from three-body dynamics (ZKL), general relativistic effects (orbit-averaged 2PN term), and tidal friction. For more complete details, we refer the reader to \citet{toonen_evolution_2016}.

Stellar evolution was performed in TRES using an interface to the population synthesis code SeBa \citep{portegies_zwart_population_1996, toonen_supernova_2012}, which employs the fitted stellar tracks from \citet{hurley_comprehensive_2000} to rapidly evolve stars. SeBa includes mass loss due to stellar winds, with different mass loss prescriptions applied at different stages of the evolution of a star. For this project, the most relevant prescriptions are summarized as follows. During the main sequence, massive star winds were implemented as Vink winds \citep{vink_new_2000, vink_mass-loss_2001} multiplied by a correction factor of $1/3$ broadly consistent with the updated prescription from \citet{bjorklund_new_2023}. During post-main-sequence evolution, the most massive stars experience mass loss attributed to a luminous blue variable (LBV) stage, modelled in SeBa following \citet{belczynski_maximum_2010} as a constant wind mass loss rate $\dot{m} = -1.5\times 10^4$ M$_\odot$/yr.

Massive stars may undergo a supernova at the end of their lives. In TRES, natal velocity kicks due to both asymmetric mass ejection, and mass loss can be included in the simulation. In this work, we set the former effect to zero. When a supernova occurred in a triple, the mean anomaly of the inner and outer orbits were sampled from a uniform distribution in $\{0, 2\pi\}$, used to calculate the positions and velocities of each body given the orbital elements at the time of the supernova. The mass lost was subtracted from the body undergoing supernova, velocity kicks were applied, and orbital elements were re-calculated using the state vectors.

We simulated a total of $52464$ triple systems: $28629$ with a metallicity of $Z=0.0005$ and $23835$ with a metallicity of $Z=0.005$. The uneven distribution of the total number of systems arises from the numerical problems encountered during the simulations of certain triples. We discarded triples that, at ZAMS, are either dynamically unstable or experience Roche-lobe overflow (RLOF). The initial stellar mass of the primary star was assumed to follow a Kroupa distribution \citep{kroupa_variation_2001} with the minimum and maximum stellar mass set to 16 M$_\odot$ and 100 M$_\odot$, respectively. The value for the lower limit comes from the minimum mass required to form a BH in SeBa. The mass of the secondary star was determined by sampling the mass ratio $q=m_1/m_2$ from a uniform distribution in $\{0, 1\}$, while simultaneously keeping the same upper and lower limits as for the primary. Similarly, the tertiary stellar mass was assumed to follow a uniform distribution in the outer mass ratio $q_\text{out} \equiv m_3/(m_1+m_2)$. For the orbital properties, the initial inner and outer eccentricities were assumed to follow a thermal distribution with limits $\{0, 0.9\}$, while the mutual inclination was determined by sampling a circular uniform distribution of $\cos{i}$ with limits $\{-1, 1 \}$. The outer semi-major axis was drawn from a uniform distribution in log-space, with limits $\log{\{0.5, 5 \times 10^6\}} \text{R}_\odot$. For the inner binary, the semi-major axis was similarly sampled from a distribution in log-space. However, the lower limit of the distribution was set to depend on the initial primary mass. This was intended to produce fewer systems that would interact during the pre-BH evolution (as the maximum radius achieved by a star during its evolution in SeBa depends strongly on its initial mass), thereby increasing the number of BH-BH systems and reducing uncertainties in our final statistics. We fitted a polynomial to the data simulated with SeBa for metallicities of $Z=0.005$ and $Z=0.0005$, giving an estimated maximum radius as a function of the initial primary mass $m_{1,i}$ as

\begin{equation}
    R_\text{max}(m_{1,i}) = p_0 + p_1m_{1,i} + p_2 m_{1,i}^2,
\end{equation}

where the coefficients $p$ are given in Table \ref{tab:rmax_coeffs}. The minimum inner semi-major axis was then calculated from the approximate Roche radius \citep{eggletonAproximationsRadiiRoche1983}:

\begin{equation}
    a_\text{in, min} = \frac{(1 + q_\text{in})^{0.2}R_\text{max}}{0.44q_\text{in}^{0.33}}.
\end{equation}

\begin{table}[]
\caption{Coefficients of polynomials fitted to the simulated data of the maximum radius as a function of initial stellar mass.}
\label{tab:rmax_coeffs}
\centering
\begin{tabular}{@{}llll@{}}
\toprule
Z        & $p_0$   & $p_1$   & $p_2$   \\ \midrule
$0.005$  & 716.906 & 29.903  & -0.0912 \\
$0.0005$ & 117.676 & 53.2623 & 0.0     \\ \bottomrule
\end{tabular}
\end{table}

The distributions of the initial conditions are based on a combination of observational constraints and extrapolations. The mass ratios and orbital separations are based on observations of massive binaries (e.g. \citet{sana_binary_2012, kobulnicky_toward_2014}), while orbital separations follow observations of both massive binaries and triples \citet{kobulnicky_new_2007, moe_mind_2017}. The thermal distribution of the eccentricities are based on observations of O/early-B-type stars as compiled by \citet{moe_mind_2017}. The distribution of mutual inclination has been shown to depend on the outer separation, with triples that have $a_\text{out} \geq 1000$au generally showing misalignment between the inner and outer orbits \citep{tokovinin_orbit_2017}. As this study focuses exclusively on triples that satisfy this criterion, a distribution of $\sin{i}$ uniform in $(-1, 1)$ is justified.

Triple systems were simulated in TRES until one of the following stopping conditions were met: (1) RLOF occurred in either the inner or outer binary; (2) the triple disintegrated (became unbound) following a supernova; (3) the triple became dynamically unstable; (4) the age of the triple reached $13.5$ Gyr; or (5) the inner binary of the triple merged due to gravitational wave emission. We did not model any stellar interaction as our focus lies on non-interacting triples. 

\subsection{Simulating black hole mergers with Syzygy}\label{sec:method_syzygy}

Triples that remained gravitationally bound following the secondary star's supernova were simulated from the point of BHB formation using a direct n-body code. This includes dynamically stable triple and triples that are dynamically unstable at BHB formation. As TRES only provides information about the triple orbital elements, these must first be converted to state vectors (positions and velocities) to run the n-body simulations. For dynamically stable systems, the initial positions and velocities were set to their values at apoapsis of the current orbital configuration. For the dynamically unstable systems, we generated ten instances of each system, varying only the initial outer true anomaly. This artificial inflation helps mitigate the uncertainties in the statistics of the final outcomes resulting from the chaotic nature of triples on the edge of stability.

In this study, we used the n-body code \texttt{Syzygy} \citep{bruenech_massive_2025}, a direct n-body integrator written in the Julia programming language. We included post-Newtonian correction terms up to an order of 2.5 to model pericentre precession, circularization, and orbital decay due to gravitational wave emission. We used the equations from \citet{blanchet_post-newtonian_2024} in the general frame. These are presented in Appendix \ref{appendix:PN_acceleration}. \texttt{Syzygy} uses the DifferentialEquations.jl \citep{rackauckas_differentialequationsjl_2017} ecosystem to solve the governing differential equations, enabling a wide selection of ODE solver algorithms. For this project, we utilized Vern9 \citep{verner_numerically_2010}: Verner's “Most Efficient” 9/8 Runge-Kutta method with a lazy ninth-order interpolant combined with absolute and relative error tolerances of $10^{-13}$. This solver is highly efficient for simulations that require extremely low tolerances and high accuracy. Small tolerances were required for these simulations due to their long integration times and extreme instantaneous accelerations resulting from orbits with very high eccentricities.

Triple systems were simulated with \texttt{Syzygy} until one of the following stopping conditions was triggered:

\begin{itemize}
    \item Direct collision between two bodies: For compact objects, the code checks whether the instantaneous separation is smaller than a multiple of their mutual gravitational radius, defined as $r_g \equiv G(m_1 + m_2)/c^2$. In these simulations, we terminated the solver if the separation $d \leq 50r_g$. 
    \item Ejection. For triple systems near or at the stability limit, strong three-body dynamics may eject one of the bodies, leaving a single star and an isolated binary that are no longer relevant to this study.
    \item Black hole merger: If the semi-major axis of the inner binary became less than or equal to $1$ R$_\odot$, we terminated the simulation and labelled the system a merger. These BH binaries had not yet merged at this stage, although the inner binary had lost enough energy through GW emission to shrink the inner semi-major axis by roughly $3-4$ orders of magnitude. As a result, these completely decoupled from the tertiary star and merged independently due to further gravitational wave emission. To save computational resources, we stopped the simulation at this point and simulated the final evolution of the semi-major axis and eccentricity using the \citep{peters_gravitational_1964} equations. See Sect. (\ref{sec:final_inspiral}).
    \item CPU time: We set a maximum CPU time of 48 hours per simulation to further limit computational resource usage.
\end{itemize}

\subsection{Modelling spin precession}

The triple systems that experienced an inner BHB merger (i.e. the systems satisfying criterion 3 of the stopping conditions) were simulated again with the inclusion of black holes spin vector precession. The spin evolution of the bodies in the inner binary was calculated by integrating the spin precession equations up to $2$PN order (see Appendix \ref{appendix:spin_precession} for the full equations). We implemented this as a callback for the n-body solver, which works in the following way. After the n-body solver advanced from $t=t_\text{prev}$ to $t=t_\text{prev} + \Delta t$, the spins were evolved by integrating the spin precession equations for a total of $\Delta t$ time units. During this interval, the positions and velocities were linearly interpolated between $t=t_\text{prev}$ and $t=t_\text{prev} + \Delta t$. We used the same numerical setup (algorithm and tolerances) for the spin integrator as for the n-body solver. We note that this method omits back-reactions on the orbits due to spin-orbit effects, which can introduce additional precession of the orbital angular momentum and eccentricity vector. However, a noted by \citet{antonini_precessional_2018}, these terms can be neglected during the ZKL phase since the orbital angular momentum of the inner binary is much greater than the spin magnitudes.

For the initial values of the BH spin vectors, we performed two sets of simulations: one in which the inner BHs begin with spin vectors aligned with the inner orbital angular momentum, i.e. $\vbs{S}_{i, 0} = Gm_i^2\chi\hbs{L}$, where $\chi$ is the dimensionless spin parameter. For the other population  the initial spin vectors are determined by randomly sampling a point on the unit sphere. For both populations, we set the dimensionless spin parameter to be $0.9$.

\subsection{Modelling the final inspiral}\label{sec:final_inspiral}

To reduce computational time and resources, we stopped the n-body simulation when the inner binary semi-major axis reached $1$ R$_\odot$ after having shrunk due to GW emission. The remaining evolution of the orbital elements can accurately be described by the secular equations from \citet{peters_gravitational_1964}:

\begin{equation}
    \frac{da}{dt}\Big|_\text{GW} = -\frac{64}{5}\frac{G^3m_1 m_2(m_1 + m_2)}{c^5 a^3(1 - e^2)^{7/2}} \left(1 + \frac{73}{24}e^2 + \frac{37}{96}e^4 \right),
\end{equation}

\begin{equation}
    \frac{d\vbs{e}}{dt}\Big|_\text{GW} = -\frac{304}{15}\frac{G^3m_1 m_2(m_1 + m_2)}{c^5 a^4(1 - e^2)^{5/2}} \left(1 + \frac{121}{304}e^2 \right)\vbs{e},
\end{equation}

\begin{equation}
    \frac{d\vbs{L}}{dt}\Big|_\text{GW} = -\frac{32}{5}\frac{G^{7/2}m_1^2 m_2^2\sqrt{m_1 + m_2}}{c^5 a^{7/2}(1 - e^2)^2} \left(1 + \frac{7}{8}e^2 \right) \hbs{L}.
\end{equation}

We also included pericentre precession as

\begin{equation}
    \frac{d\vbs{e}}{dt}\Big|_\text{pr} = -\frac{3G(m_1 + m_2)}{c^2 a(1 - e^2)}\sqrt{\frac{G(m_1 + m_2)}{a^3}} \hbs{L} \times \vbs{e}.
\end{equation}

To obtain the final spin evolution, we also solved the orbit-averaged spin precession equations from \citet{barker_gravitational_1975}. The total change in the spin of body $1$ is given by

\begin{equation}
    \frac{d\vbs{S}_1}{dt} = (\vbs{\Omega}_{1, \rm{dS}} + \vbs{\Omega}_{1, \rm{LT}} + \vbs{\Omega}_{1, \rm{QM}}) \times \vbs{S}_1,
\end{equation}

where

\begin{equation}
    \vbs{\Omega}_{1, \rm{dS}} = \frac{G}{2c^2}\frac{(4 + 3q)\hbs{L}}{a^3 \sqrt{(1 - e^2)^3}},
\end{equation}

\begin{equation}
    \vbs{\Omega}_{1, \rm{LT}} = \frac{G}{2c^2}\frac{S_2}{a^3 \sqrt{(1 - e^2)^3}} \left[\hbs{S}_2 - 3 (\hbs{L} \cdot \hbs{S}_2) \hbs{L} \right],
\end{equation}

\begin{equation}
    \vbs{\Omega}_{1, \rm{QM}} = \frac{G}{2c^2}\frac{qS_2}{a^3 \sqrt{(1 - e^2)^3}} \left[\hbs{S}_1 - 3 (\hbs{L} \cdot \hbs{S}_1) \hbs{L} \right].
\end{equation}

Here, $q \equiv m_2/m_1$ is the mass ratio and $\vbs{S}_1$ and $\vbs{S}_2$ denote the spin-vectors of bodies $1$ and $2$. The equivalent equations for $\vbs{S}_2$ were obtained by swapping the labels $1 \Rightarrow 2$ (and, consequently, $q \Rightarrow q^{-1}$). Compact object spins also introduce changes to the orbital elements as back-reactions. The change in the orbital angular momentum and eccentricity due to spin-orbit back-reactions are given by

\begin{equation}
\begin{aligned}
    \frac{d\vbs{x}}{dt} \Big|_{\text{SO}} = (&\vbs{\Omega}_{1, \rm{dS}, \rm{SO}} + \vbs{\Omega}_{2, \rm{dS}, \rm{SO}} + \vbs{\Omega}_{\rm{LT}, \rm{SO}} + \\ &\vbs{\Omega}_{1, \rm{QM}, \rm{SO}} + \vbs{\Omega}_{2, \rm{QM}, \rm{SO}}) \times \vbs{x},
\end{aligned}
\end{equation}

where $\vbs{x} \equiv \{\vbs{L}, \vbs{e}\}$, and

\begin{equation}
    \vbs{\Omega}_{1, \rm{dS}, \rm{SO}} = \frac{G}{2c^2}\frac{S_1(4 + 3q)L}{a^3 \sqrt{(1 - e^2)^3}}\left[\hbs{S}_1 - 3 (\hbs{L} \cdot \hbs{S}_1) \hbs{L} \right],
\end{equation}

\begin{equation}
\begin{aligned}
    \vbs{\Omega}_{\rm{LT}, \rm{SO}} &= \frac{G}{2c^2}\frac{3S_1S_2}{a^3 L\sqrt{(1 - e^2)^3}} \Bigl[ (\hbs{L} \cdot \hbs{S}_1) \hbs{S}_2 + (\hbs{L} \cdot \hbs{S})\\
    &+ \left\{ (\hbs{S}_1 \cdot \hbs{S}_2) - 5(\hbs{L} \cdot \hbs{S}_1) (\hbs{L} \cdot \hbs{S}_2) \right\} \hbs{L} \Bigr],
\end{aligned}
\end{equation}

\begin{equation}
    \vbs{\Omega}_{1, \rm{QM}, \rm{SO}} = \frac{G}{2c^2}\frac{3qS_1^2}{2a^3L \sqrt{(1 - e^2)^3}} \left[ 2(\hbs{L} \cdot \hbs{S}_1) \hbs{S}_1 + \left(1 - 5(\hbs{L} \cdot \hbs{S}_1)^2 \right) \right].
\end{equation}

The full expression for the change in $\vbs{L}$ and $\vbs{e}$ due to both gravitational radiation, precession, and spin-orbit back-reaction is thus

\begin{equation}
    \frac{d\vbs{L}}{dt} = \frac{d\vbs{L}}{dt}\Big|_\text{GW} + \frac{d\vbs{L}}{dt}\Big|_\text{SO},
\end{equation}

\begin{equation}
    \frac{d\vbs{e}}{dt} = \frac{d\vbs{e}}{dt}\Big|_\text{GW} + \frac{d\vbs{e}}{dt}\Big|_\text{pr} + \frac{d\vbs{e}}{dt}\Big|_\text{SO}.
\end{equation}

where $\vbs{S}_1$ is the spin-vector of body $1$, $q \equiv m_2/m_1$ is the mass ratio of the two inspiraling bodies, and $\vbs{L}$ is the orbital angular momentum. The evolution of the spin-vector of body $2$ was obtained by swapping the subscript labels $1 \Leftrightarrow 2$ (and, consequently, $q \Rightarrow q^{-1}$).

Using the orbital properties of the merging systems at the time of n-body termination, we solved the above equations to obtain the final evolution of the semi-major axis, eccentricity, angular momentum, and spins. We solved the system of secular equations using the Tsitouras 5/4 method with absolute and relative error tolerances of $10^{-6}$. We stopped the solver when either the semi-major axis equalled $10r_g$, at which point the secular equations were no longer valid, or when the system reached a peak gravitational wave frequency of $10$ Hz, i.e. when \citep{wen_eccentricity_2003}  

\begin{equation}\label{eq:fGW}
f_\text{GW} \equiv \frac{\sqrt{G(m_1+m_2)}}{\pi(1+e)^{1.195}a(1 - e^2)^{3/2}} \geq 10 \, \text{Hz}.
\end{equation}

\section{Simulation results}\label{sec:results}

\subsection{Triple population synthesis}

The vast majority of the triple systems we simulated ended up experiencing either disintegration due to one or more supernovae or stellar interaction in the form of RLOF before the two inner components formed BHs (Table \ref{tab:tres_outcomes}). Around $0.5\%$ of the triples remain bound but become dynamically unstable following the formation of an inner BHB, while $\sim 1\%$ result in stable triples with inner BHBs. In absolute numbers, for $Z=0.0005$ ($Z=0.005$), this corresponds to $562$ ($335$) and $1077$ ($850$) systems, respectively. None of the triple systems ends up with an inner BHB merger in TRES, meaning that all the stable BHBs reach the Hubble time without merging. 

\begin{figure}
    \centering
    \includegraphics[width=1.0\linewidth]{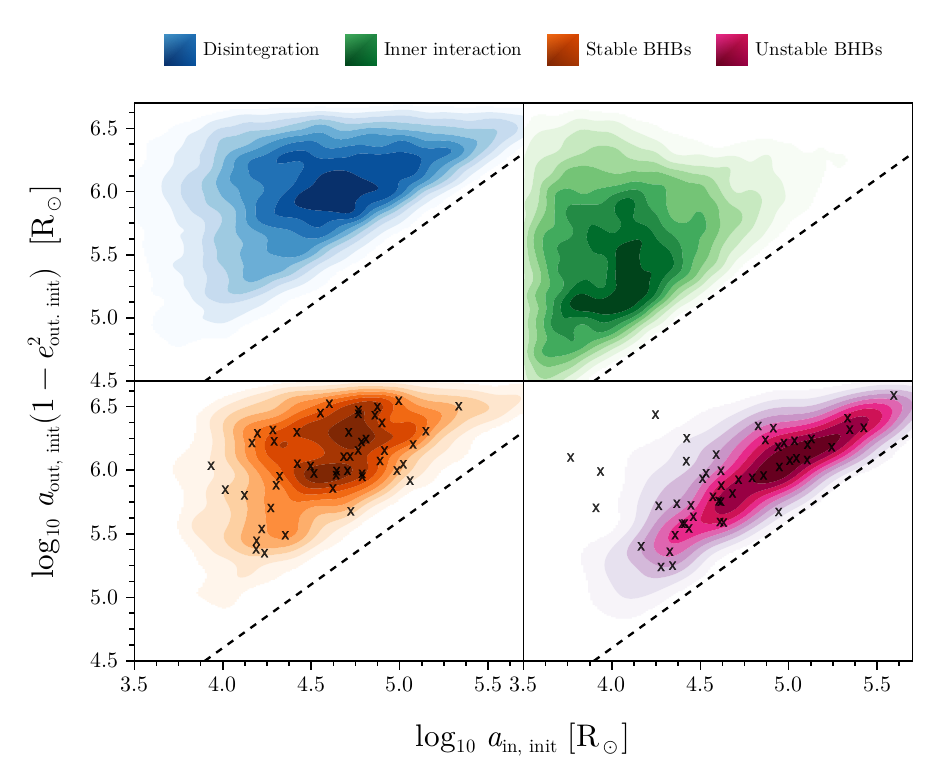}  
    \caption{Kernel density plots of the initial outer semi-latus rectum as a function of the initial inner semi-major axis for various outcomes of the triple population synthesis simulations. Disintegration refers to triples that become unbound due to mass loss from one or more supernovae; inner interaction denotes systems that experienced inner or outer RLOF; and stable and unstable BHBs are systems that avoided stellar interactions and survived both inner supernovae, forming either a stable or dynamically unstable triple with an inner BHB. The dashed line shows where $a_\text{out}(1 - e_\text{out}^2) = 5a_\text{in}$, approximating the limit of dynamical instability defined by Eq. (\ref{eq:mardling_aarseth}). The crosses indicate the ZAMS properties of triples that eventually ended up in a merger.}
    \label{fig:TRES_initial_smas}
\end{figure}

Even with an initial distribution of wide orbits, stellar interactions are difficult to avoid due to two main mechanisms: the mass-dependent expansion of stellar radii during the post-MS evolution of massive stars and the large eccentricities in the inner binary due to ZKL. Interactions can be triggered by either process individually or by both. Stellar interactions occur predominantly in systems with initially smaller inner separations (Fig \ref{fig:TRES_initial_smas}), and consequently the majority of systems that form BHBs have initial inner semi-major axes $\geq 10^4$ R$_\odot$. This result highlights why wide triples were initially chosen as the focus of this study.  While a larger inner separation makes it less likely for a triple system to experience stellar interaction in the inner binary, it also makes it easier for the system to disintegrate when a supernova occurs. This simply results from the lower gravitational binding energy combined with explosive mass loss, which itself is directly related to stellar mass loss during a supernova. In SeBa, the mass of the remnant BH is determined by three main effects: the mass of the C-O core of the progenitor star at the time of BH formation, the amount of fallback on the remnant as described by \citet{fryer_compact_2012}, and finally the amount of mass lost due to neutrinos. Following \citet{belczynski_first_2016}, the fraction of mass lost due to neutrinos is $10$\% of the proto-BH mass (as determined by the C-O core mass) plus the fallback mass. As noted by \citet{dorozsmai_hierarchical_2025}, the process behind neutrino mass loss is highly uncertain, with different stellar evolution codes employing different prescriptions and models. Since we did not include natal velocity kicks in our simulations, the total amount of kick imparted onto a black hole following the supernova arises from the total mass lost (gas plus neutrinos). For the most massive stars, most of the mass loss is due to neutrinos, as these stars experience near-complete fallback. A decrease in the neutrino mass loss (see \citet{dorozsmai_hierarchical_2025}) could potentially lead to an increase in the number of triples that survive the two supernovae required to form the inner BHB. This effect could be particularly efficient given the wide orbits studied in this work, as they are inherently more susceptible to becoming unbound due to mass loss from the weaker gravitational binding energy.

\begin{table*}[]
\caption{Result of population synthesis simulations with TRES.}
\label{tab:tres_outcomes}
\centering
\begin{tabular}{lrllllll}
\hline 
Metallicity                      & \multicolumn{1}{l}{N simulated}         & \begin{tabular}[c]{@{}l@{}}Fraction\\ population\end{tabular}             & \begin{tabular}[c]{@{}l@{}}Fraction\\ disrupted\end{tabular} & \begin{tabular}[c]{@{}l@{}}Fraction\\ interacted\end{tabular} & \begin{tabular}[c]{@{}l@{}}Fraction\\ unstable\end{tabular} & \begin{tabular}[c]{@{}l@{}}Fraction \\ unstable BHBs\end{tabular} & \begin{tabular}[c]{@{}l@{}}Fraction \\ stable BHBs\end{tabular} \\ \hline
\multicolumn{1}{l|}{$Z=0.0005$}  & \cellcolor[HTML]{FFFFFF}{\color[HTML]{000000} 28629} & \cellcolor[HTML]{FFFFFF}{\color[HTML]{000000} 0.28} & \cellcolor[HTML]{FFFFFF}{\color[HTML]{000000} 0.53}           & \cellcolor[HTML]{FFFFFF}{\color[HTML]{000000} 0.31}           & \cellcolor[HTML]{FFFFFF}{\color[HTML]{000000} 0.10}         & 0.02                                                              & \cellcolor[HTML]{FFFFFF}{\color[HTML]{000000} 0.04}             \\
\multicolumn{1}{l|}{$Z=0.005$} & \cellcolor[HTML]{FFFFFF}{\color[HTML]{000000} 23835} & \cellcolor[HTML]{FFFFFF}{\color[HTML]{000000} 0.28} & \cellcolor[HTML]{FFFFFF}{\color[HTML]{000000} 0.50}           & \cellcolor[HTML]{FFFFFF}{\color[HTML]{000000} 0.37}           & \cellcolor[HTML]{FFFFFF}{\color[HTML]{000000} 0.08}         & 0.01                                                              & \cellcolor[HTML]{FFFFFF}{\color[HTML]{000000} 0.04}            
\end{tabular}
\tablefoot{The first two columns show the value of the metallicity used when modelling the stellar evolution and the total numbers of systems simulated for each metallicity. The third column gives the fraction of the full massive triple population that our subpopulations represent. The next columns describe the fraction of the simulated triples in which a body in either the outer or inner binary become unbound, the fraction of triples that experienced any form of stellar evolution, such as RLOF, and the fractions of triples that (1) destabilized, (2) destabilized with an inner BHB, and (3) formed an inner BHB and remained dynamically stable.}
\end{table*}

\subsection{Mergers only occur in n-body simulations}\label{subsec:nbody_discussion}

The $897$ dynamically unstable and $1927$ stable systems were simulated from the point of BHB formation in \texttt{Syzygy}. We identify a total of $50$ dynamically stable triples that experienced a merger in the inner binary, and $135$ mergers from the unstable systems. This corresponds to about $0.03 \%$ and $0.008\%$ of the total massive triple population, respectively. As reported, the same systems that merged in the n-body simulations did not end up with the same outcome when they were evolved using the orbit-averaged equations used by \texttt{TRES}. There are two possible reasons for this. First, the region of parameter space occupied by the merging triples may be poorly approximated by the orbit-averaged equations, meaning that certain dynamical properties (e.g. extreme eccentricities) do not emerge in the secular code. Second, the orbit-averaged equations for the general relativistic effects from \citet{peters_gravitational_1964} may also be poor approximations in the very-high-eccentricity regime. In fact, both are likely to be true, as found, for example, by \citet{antonini_binary_2017} and \citet{silsbee_lidov-kozai_2017}. The consequence of the latter is that, while the Peters equations give a slightly different evolution of the orbital elements at very high eccentricities, the difference ultimately effects the merger time, with the orbit-averaged equations predicting a later merger time compared to n-body with post-Newtonian terms. We find that an example binary system consisting of $m_1,m_2 = 23,20$ M$_\odot$ with a periapsis of $0.1$R$_\odot$ and an eccentricity of $0.9999$ has a final merger time approximately $16\%$ shorter in an n-body simulation than that predicted by the Peters equations. This may therefore explain the lack of mergers in the secular simulations, as systems did not merge within the Hubble time due to longer orbital shrinkage timescales caused by high eccentricities combined with the Peters equations.

The other effect is the loss of accuracy in the orbit-averaged equations for the three-body dynamics when the triple enters a specific part of parameter space. This is often called the semi-secular regime \citet{katz_rate_2012, antonini_black_2014} and is defined as the region where the angular momentum of the inner binary can change by order of itself over a single outer orbital period. This holds if

\begin{equation}\label{eq:semisec}
    \sqrt{1 - e} \lessapprox 5\pi q_\text{out} \left(\frac{a_\text{in}}{a_\text{out}(1 - e_\text{out})} \right)^3.
\end{equation}

\begin{figure}
    \centering
    \includegraphics[width=1.0\linewidth]{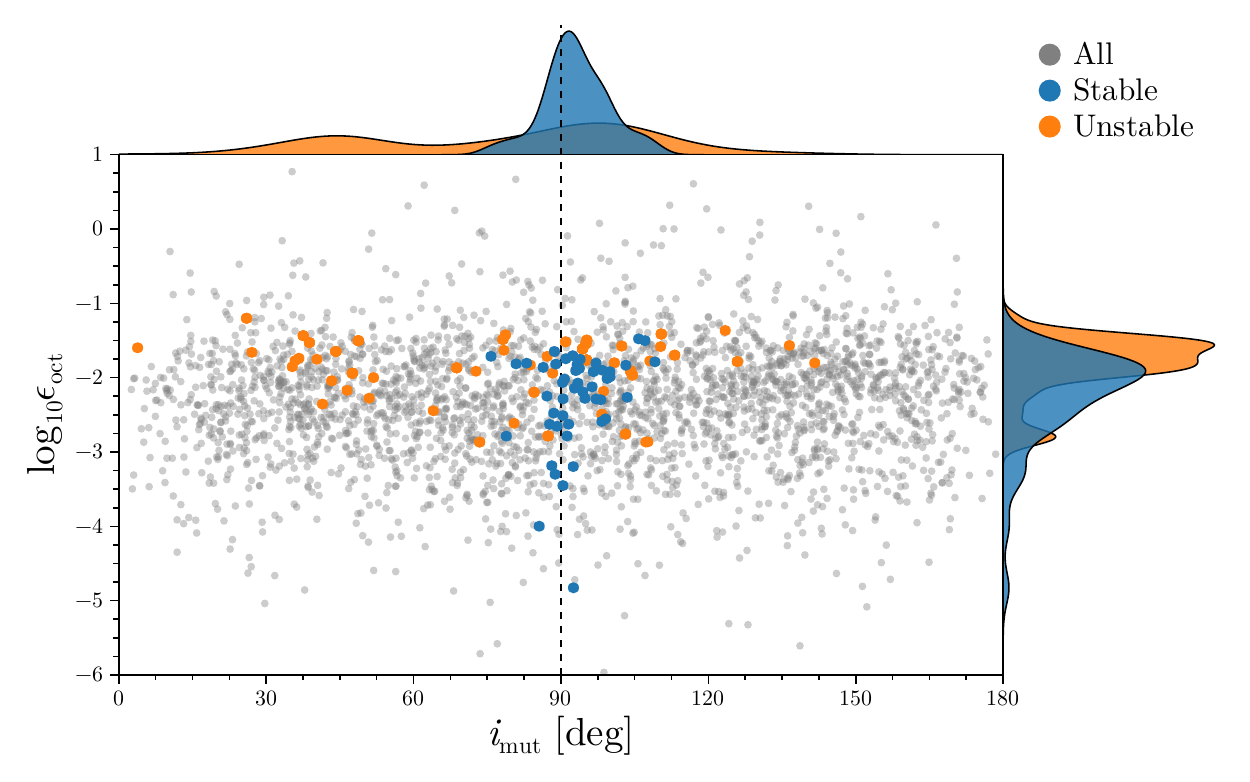}
    \caption{Octupole parameter as a function of the mutual inclination at the time of BHB formation. The grey dots show all the systems that did not result in a merger, while blue and orange represent mergers from dynamically stable and unstable populations, respectively.}
    \label{fig:epsoct_vs_imut}
\end{figure}

In this regime, the triple remains hierarchical and not on the limit of stability as defined by Eq. (\ref{eq:mardling_aarseth}). Nonetheless, the changes to the inner binary due to the tertiary are on timescales short enough such that certain dynamics are excluded when using the orbit-averaged equations. More specifically, as noted by \citet{naoz_eccentric_2016}, the maximum eccentricity of the inner binary is greater than the predicted value from the orbit-averaged equations if specific angular momentum becomes smaller than the right-hand side of Eq. (\ref{eq:semisec}). To quantify this effect, we plot the value of the octupole parameter as a function of the mutual inclination at the point of BHB formation in Fig. (\ref{fig:epsoct_vs_imut}) for both populations, with the inclusion of BHB systems that did not merge. The majority of stable triples that produce merging binaries have octupole parameters in $(10^{-3}, 10^{-2})$ and inclinations close to $90^\circ$, while for higher values of $\epsilon_\text{oct}$, merging systems have inclinations further from $90^\circ$. Moreover, the unstable systems that merged have a substantially wider range of inclinations, with two main clusters around $40^\circ$ and $100^\circ$. The consequences of these results can be summarized as follows: an inclination close to 90$^\circ$ is necessary to achieve high eccentricities required for merger; however, this dependence becomes slightly weaker if the octupole term is high enough. Additionally, systems near the stability threshold have much weaker dependence on the mutual inclination, enabling mergers to occur even when $i_\text{mut}$ is far from $90^\circ$.

\subsection{Properties required to form a BHB merger}

\begin{figure*}
\begin{tabular}[t]{@{}cc@{}} \\[-\dp\strutbox]
    \subfloat{%
        \includegraphics[width=12cm]{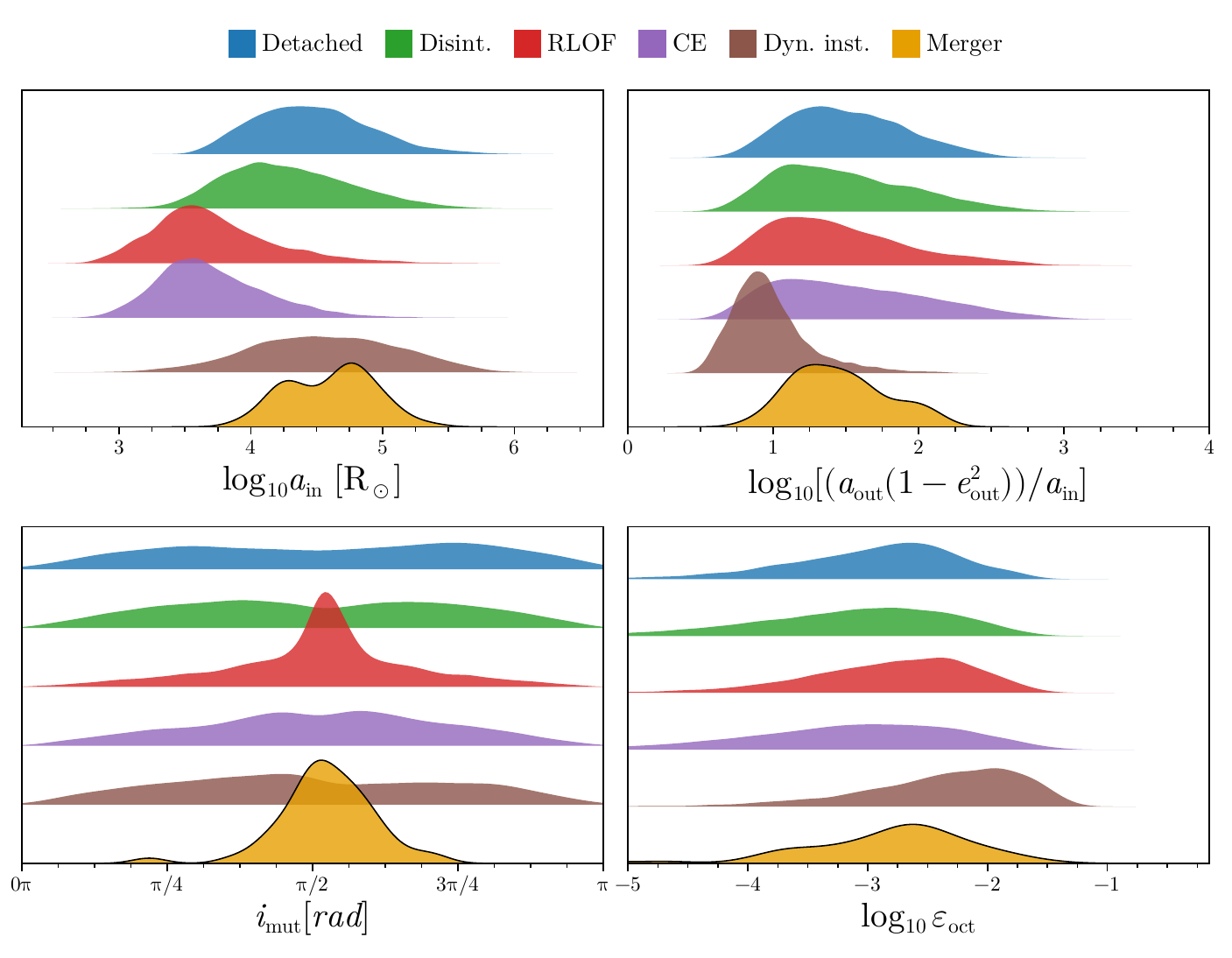}
    }\\
    \subfloat{%
        \includegraphics[width=12cm]{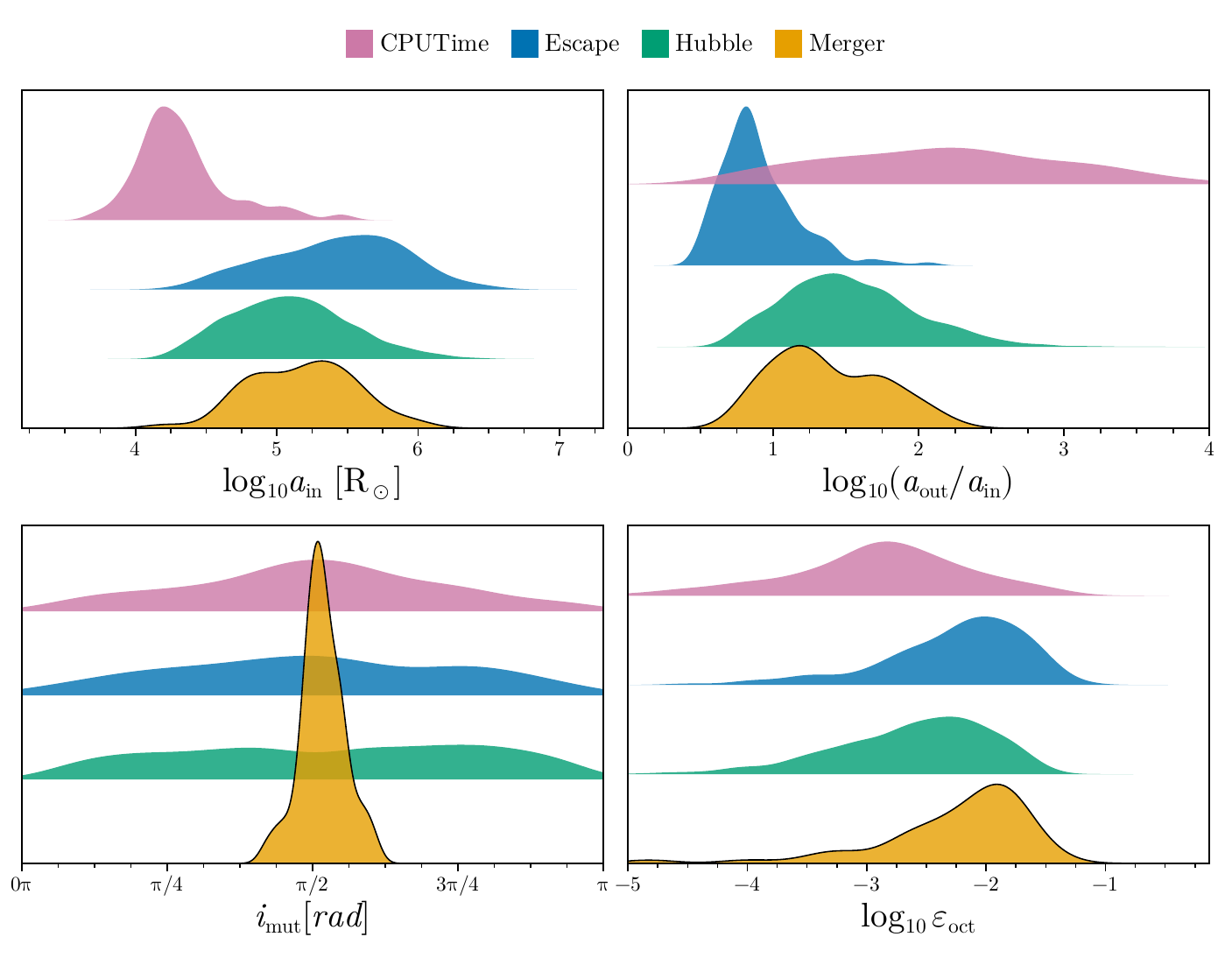}
    }
\end{tabular}\hfill
\begin{minipage}[t]{0.3\textwidth}
    \caption{Comparing distributions of orbital properties at various points in the evolution of massive triples. Top: ZAMS properties of systems that experience different outcomes in TRES, compared with that of the systems that eventually end up as a merger. Bottom: Properties of systems at the onset of BHB formation, with different colours indicating different outcomes in the n-body simulation.}
    \label{fig:properties_at_ZAMS_and_BHBonset}
\end{minipage}
\end{figure*}

To understand which initial orbital properties lead to different outcomes, we plot selected parameters at two evolutionary stages (Fig. \ref{fig:properties_at_ZAMS_and_BHBonset}): at system birth (ZAMS) (upper panel) and at BHB formation (lower panel). Triple systems that ultimately merge have distinct properties compared to those that experience other outcomes, with the precondition being the initial inclination between the inner and outer orbits. Only systems with $i_\text{mut, init} \in \{40, 130 \} ^\circ$ result in a BHB merger, even though the initial distribution of inclinations for all triples was taken to be uniform. This is not a surprising result, as the high eccentricities required to shrink the wide inner orbits can only be achieved with inclinations close to $90^\circ$ (Sect. \ref{sec:primer}; Fig. \ref{fig:epsoct_vs_imut}). However, a near-perpendicular orbital configuration does not mean that the system will definitely result in a merger, as this configuration also causes strong ZKL oscillations before BHB formation. In the upper panel of Fig. \ref{fig:properties_at_ZAMS_and_BHBonset}, we see a strong peak in the mutual inclination around $90^\circ$ for the systems that experience stable mass transfer (RLOF). These are tight-orbit systems with a closer tertiary companion that enable the inner binary to reach high eccentricities early in its evolution. This highlights the need for the triple system's parameters to occupy a very specific region of the parameter space: the inclination must be close to $90^\circ$ to induce strong dynamics, while the orbital ratios must ensure that the oscillation timescales are not shorter than the evolutionary timescales of the BH progenitors. Without taking pre-BH stellar evolution into account, generating a triple with an inner BHB that would not undergo some form of stellar interaction or disintegration before BH formation is therefore highly non-trivial.

At the time of BHB formation, the inner orbits of the merging systems are wider than at the ZAMS due to the mass lost from supernovae (Fig. \ref{fig:properties_at_ZAMS_and_BHBonset}, lower panel). As a direct consequence, the ratio of the outer semi-latus rectum to the inner semi-major axis decreases slightly, which directly precipitates shorter ZKL timescales. The mutual inclination is even more strongly peaked around $90^\circ$, indicating that the merging triples changed their mutual inclination between ZAMS and BHB formation. The octupole term is now substantially higher for most of the systems than at ZAMS, with an average value near $10^{-2}$, compared to approximately $10^{-3}$ at ZAMs. This higher octupole value is required to reach the extreme eccentricities necessary to merge the inner binary, and it is clear that this value is not present at ZAMS, but is instead induced by stellar evolution toward BH formation, with mass loss during the supernova being the dominant contributor.

Compared to the other outcomes from the n-body simulations, we see that the CPU time systems have a unique combination of parameters: they have tight inner binaries, wide outer orbits, a broad inclination distribution centred around $90^\circ$, and a wide distribution of octupole values with the majority having $\varepsilon_\text{oct} \in \{10^{-3}, 10^{-2}\}$. Triples with tight inner binaries and wide outer orbits took longer to simulate due to the small time steps required to resolve the inner binary. At the same time, large values of $a_\text{out}/a_\text{in}$ results in a long ZKL timescale, so these systems not only take longer to finish in CPU time, but they also take longer in their physical time before ZKL oscillations kick in. We identify a total of $332$ CPU time systems, almost twice the number of merging triples ($N=185$). An important question is whether we are missing any mergers from this pool of CPU time systems. If we consider the subset of these systems that overlap with the parameters of most merging systems\footnote{We checked for CPU time systems with $\log_{10}a_\text{in} \in (4.7, 5.5)$, $\log_{10}(a_\text{out}/a_\text{in}) \in (1, 2)$, $i_\text{mut} \in (\frac{\pi}{8}, \frac{3\pi}{8})$, and $\log_{10}\varepsilon_\text{oct} \in (-2.5, -1)$.}, we find that only eight systems fall into this overlapping region of the parameter space. This suggests that we are likely not excluding a substantial amount of mergers.

\subsection{Final merger properties}

\begin{figure}

    \centering
    \includegraphics[width=0.9\linewidth]{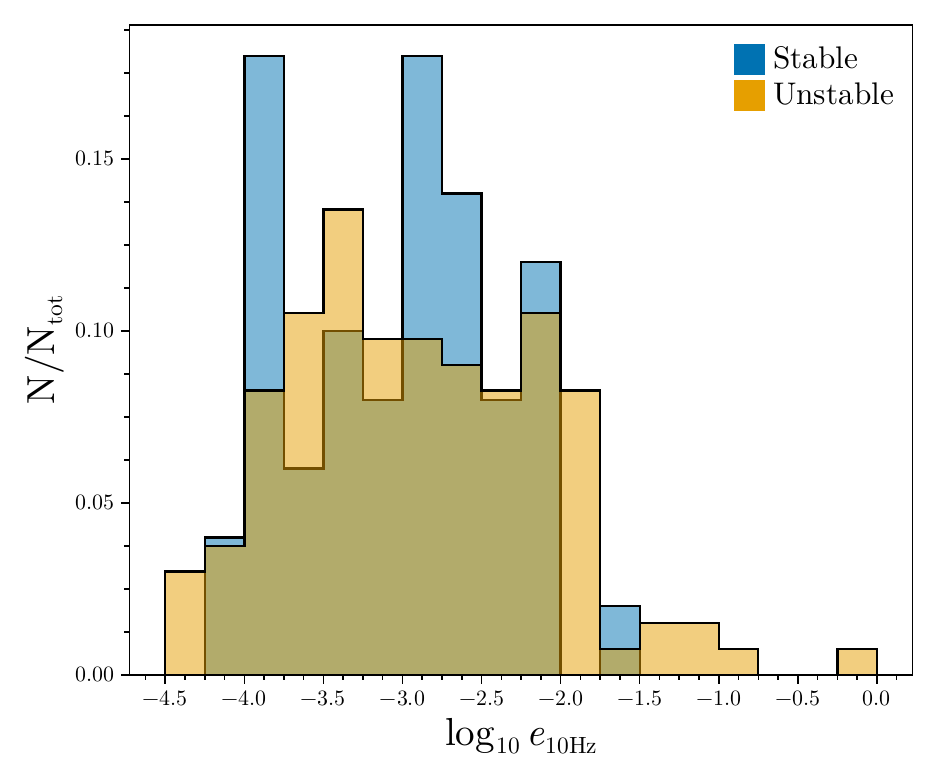}
    \caption{Distribution of eccentricities for the merging BHBs at the point when they enter the 10Hz LIGO frequency band. The blue histogram shows the eccentricities for the dynamically stable mergers, and the yellow histogram shows the mergers in triples that were dynamically unstable at BHB formation.}
    \label{fig:eLIGO_distribution}
\end{figure}

The distributions of eccentricities of the merging systems at $f_\text{GW} = 10$Hz are shown in Fig. (\ref{fig:eLIGO_distribution}). We find that $94$\% of the mergers from the dynamically stable channel fall in the range $10^{-4} \geq e_\text{10 Hz} \geq 10^{-2}$, with two prominent peaks at $e_\text{10 Hz} \approx 10^{-4}$ and $\approx 10^{-2.75}$. The systems that were dynamically unstable at BHB formation show a more uniform distribution in the same limits but with a small tail towards higher eccentricities, with about $4.5$ \% having $e_\text{10 Hz} \geq 10^{-1.5}$. The most extreme eccentricities for both the stable and unstable population were recorded to be $0.02$ and $0.84$, respectively. These distributions show that the triple-induced merging BHBs are generally more eccentric than expected from mergers in isolated binary evolution, where the binary orbit is expected to have fully circularized by the time the system enters the $10$ Hz frequency band \citep{peters_gravitational_1964}. However, the merging binaries still show a significant spread in eccentricity, with a nearly uniform distribution from $log_{10}{e_\text{10Hz}} \approx -4$ to $log_{10}{e_\text{10Hz}} \approx -2$. To determine which parameters influence the eccentricity of the merging binary upon entering the LIGO band, we calculated the Spearman rank correlation coefficient\footnote{Spearman's rank correlation coefficient quantifies the relationship between variables and can determine correlation even when the data are not linearly dependent. A value of 1 indicates a monotonical relation, while -1 means the data are inversely proportional.} for $e_\text{10Hz}$ as a function of $a_\text{in}$, $a_\text{out}$, $a_\text{out}/a_\text{in}$, and $e_\text{out}$ at the onset of BHB formation (Table \ref{tab:spearman}). We also checked correlations for $a_\text{out}/a_\text{in}$ and $\frac{a_\text{out}(1 - e_\text{out})}{a_\text{in}}$; however, the correlations were not as significant. We find that for the dynamically stable systems, the outer semi-major axis shows the strongest correlation at $-0.57$, followed by the semi-major axis ratio at $-0.32$ and inner semi-major-axis at $-0.29$, and outer eccentricity at $-0.26$. Interestingly, the unstable systems show the same general trend, but the values for the correlation coefficients are higher at $-0.7$, $-0.39$, $-0.45$, and $-0.34$, respectively. This is likely due to the proximity of the tertiary to the inner binary, which allows it to affect the inner binary in ways that extend beyond classical (eccentric) ZKL. Nevertheless, the results show that the most eccentric systems have tertiaries that are closer to the inner binary. With a proximate tertiary, each passing can induce stronger perturbations in the inner binary. This means that in just a few outer orbital periods, the inner eccentricity can increase from a value with a corresponding periapsis without substantial GW emission, to one where GWs rapidly extract energy and shrink the orbit. In the case of a more distant tertiary, the inner binary is more likely to gradually approach the periapsis where GWs become significant, resulting in an eccentricity just high enough for emission. In contrast, with a closer tertiary, the eccentricity may overshoot the critical value at which GWs extract significant amounts of energy. The inverse correlation between inner binary separation and final eccentricity indicates that additional eccentric mergers could occur if the maximum radius of the progenitor stars were smaller, enabling the formation of more compact systems that produce BHBs while avoiding stellar interactions.

\begin{table}[]
\caption{Spearman rank correlation coefficient of orbital properties at BHB formation and eccentricity at $10$Hz.}
\label{tab:spearman}
\centering
\begin{tabular}{lllll}
\hline
Population & $a_\text{in, init}$ & $a_\text{out, init}$ & $a_\text{out}/a_\text{in}$ & $e_\text{out, init}$  \\ \hline
Stable     & -0.29               & -0.57                & -0.32               & -0.26                                                                                \\
Unstable   & -0.45               & -0.7                 & -0.39               & -0.34                                                                                \\ \hline
\end{tabular}

\end{table}

\subsection{Spin evolution and final effective spin}

\begin{figure*}[b!]
    \centering
    \includegraphics[width=0.9\linewidth]{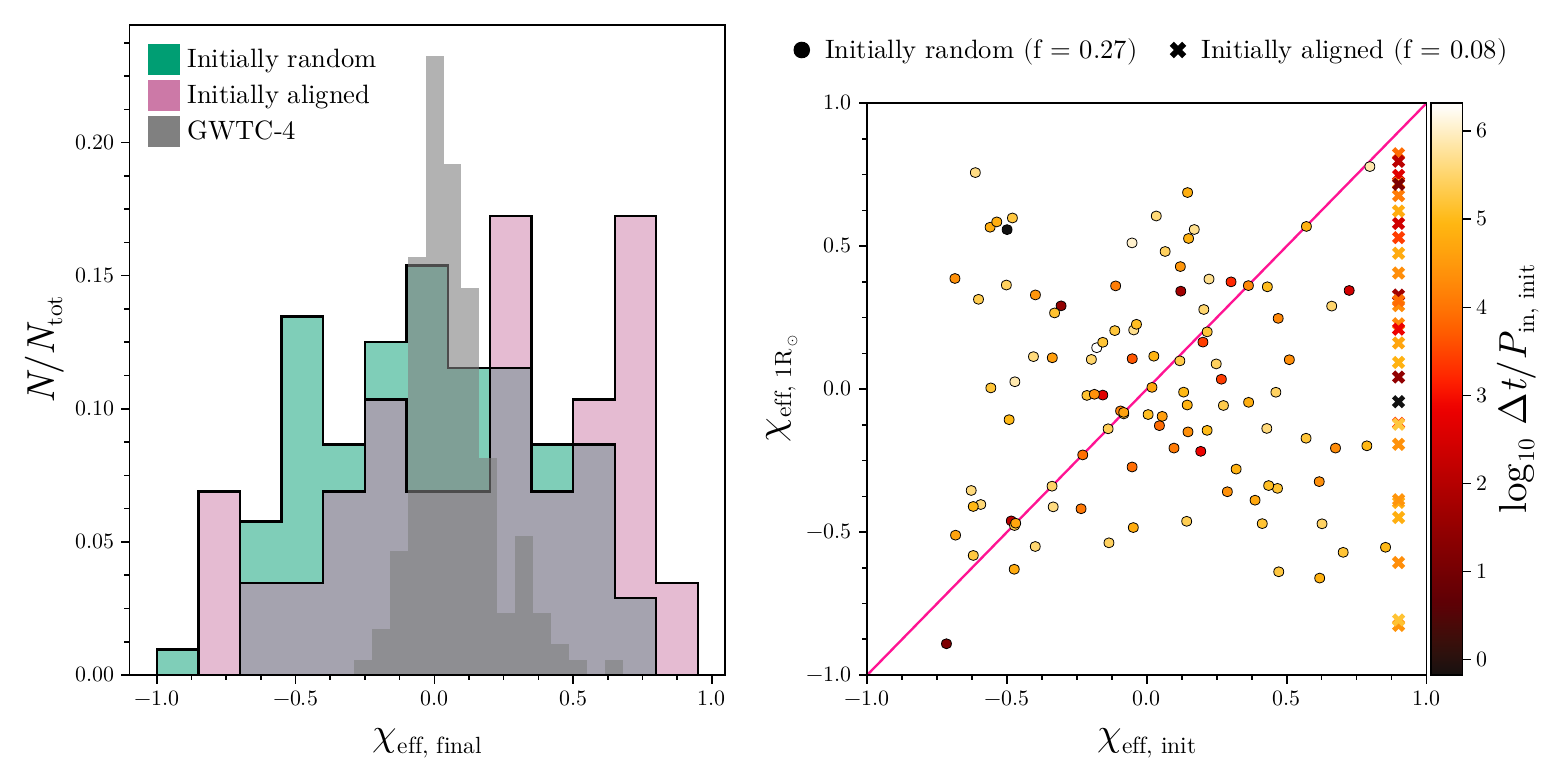}
    \caption{Overview of effective spins at different times in the evolution of merging binaries. Right: Effective spins at the time of n-body termination (when the inner binary shrinks to a separation of 1 R$_\odot$) as a function of the effective spin at the time of BHB formation. The circles show the system that had spins with an initially random orientation, while the crosses shows the systems with spins initially aligned with the inner binary angular momentum. The colour indicates the time it took for the binary to merge in units of the initial inner period. We also indicate the fraction of the total merging inner binaries that actually merged when spins were included with the $f$ parameter. Left: Distribution of the effective spins when they enter the $10$Hz LIGO band. The blue and yellow figures shows the distributions for the initially random and initially aligned spin populations, respectively, while the grey histogram shows the cumulative data from GWTC-4.0.}
    \label{fig:chi_eff_final}
\end{figure*}

The triples containing merging inner binaries were re-simulated in the n-body code with the inclusion of spin precession. After stopping the n-body code, we evolved the final inspiral of the inner binary using the secular orbit-averaged equations, this time including the secular spin terms. The left plot in Fig. (\ref{fig:chi_eff_final}) shows the final distribution of the effective spin, while the right plot shows the effective spin at the point of n-body termination as a function of the initial (at BH formation) effective spin. During the ZKL phase of the evolution, the individual spins of the black holes do not precess, meaning that their angle with respect to their initial orientation ($\angle \vec{L}(t >0),\vec{L}(t=0)$) remains roughly zero. At the same time, the orbital angular momentum vector of the inner binary rapidly precesses due to strong ZKL oscillations, with some systems experiencing angular momentum precession of the order of $\pi$ radians (see Fig. \ref{fig:example_L_angles} for a few examples). As the effective spin is by definition the dot product of the individual spins with the binary angular momentum, the effective spin of the inner binary essentially follows the precession of $L_\text{in}$. For the most extreme cases, the value of $\chi_\text{eff}$ can vary between $\sim -0.9$ and $\sim 0.9$ during the ZKL phase. Once the periapsis is small enough for GWs to dissipate significant orbital energy, the inner binary rapidly shrinks and decouples from the tertiary\footnote{By 'decouple', we mean that the timescales for ZKL becomes substantially longer than the merger time due to the increase in the semi-major axis ratio as the inner orbit shrinks.}. During the phase in which the inner binary shrinks from GW emission, the individual spin vectors precess rapidly around $\bs{L}_\text{in}$. Nevertheless, the effective spin remains constant. The consequence of this is that the effective spin at the time of merger is effectively the same as when the inner binary achieves its highest eccentricity. As a result, the final distribution of effective spins shows a broad curve that is symmetric around zero (Fig. \ref{fig:chi_eff_final}), indicating a uniform distribution \footnote{A uniform distribution of vectors on a sphere gives a peak in the angle distribution between the vectors and the unit vector in the z direction.}. This result holds for both cases of the initial spin vector angles aligned with $\vec{L}_\text{in}$ and randomly distributed. In summary, our results indicate that BHBs that merge through this channel can produce a wide range of effective spin values even when the black holes are rapidly spinning.

Due to the substantially higher computational cost of including the direct spin equations, only about $8$ \% ($N=29$) of the total systems reach the point of merger when the spins are evolved, even with a CPU time limit set to twice that of the simulations without spins. For the triples with initially randomly distributed spins, this percentage was substantially higher at $27$ \% (104 systems). As a consequence, we may be missing systems that, if evolved to merger, would exhibit a different effective spin evolution compared to the systems we fully simulated. However, we believe this is unlikely for a few reasons. Primarily, all the triples that produce mergers undergo strong ZKL cycles, which causes rapid precession of the inner angular momentum. This occurs even when the periapsis of the inner orbit is too large for the spins to undergo significant changes, meaning that the final spin-orbit misalignment is determined almost uniquely by the ZKL oscillations. In fact, $50\%$ of the merging inner binaries exhibit an inner angular momentum vector change of $ \geq 90^\circ$ at the moment of n-body termination. Second, the systems that do not finish evolving do not occupy a specific area in the parameter space of the orbit, stellar properties, or spin angles; therefore, we do not believe there is any reason these systems would evolve in a different way compared to the merging triples. Nevertheless, future research should explore methods to evolve these systems all the way to merger to ensure that the properties of the spin evolution during the ZKL phase indeed hold for the full population.

\section{Discussion}\label{sec:discussion}

\subsection{Merger rates and model uncertainties}

We followed the same procedure as \citet{dorozsmai_stellar_2024} for calculating the rate of mergers from the wide triple channel. The resulting formation efficiency and merger rate is given in Table (\ref{tab:merger_rates}). For our estimates, we assumed that the fraction of stellar systems that reside in triples, $f_\text{triple}$, is $0.73$ with the single and binary star fraction at $0.06$ and $0.21$, respectively. We did not account for higher-order multiples. While observations have shown that the multiplicity fraction depends on the mass of the primary \citep{moe_mind_2017, offner_origin_2023}, with higher-mass stars having, on average, more stellar companions, we assumed for simplicity a fixed triple fraction. We find a merger rate of $5$ for dynamically stable triples and $\sim 1.5$ for the systems that were dynamically unstable at the time of BHB formation. The latest estimates for BHB mergers from GWTC-4.0 gives $14$ - $26$ Gpc$^{-3}$ yr$^{-1}$, meaning that the triple channel can be a substantial contributor to the overall merging BHB population. These values are consistent with results from previous studies of the same channel, which range from $1 - 5$ Gpc$^{-3}$ yr$^{-1}$ (e.g. \citet{antonini_binary_2017, silsbee_lidov-kozai_2017, rodriguez_triple_2018, martinez_mass_2022}). More specifically, \citet{antonini_binary_2017} report a merger rate density of $0.3$ to $1.3$ Gpc$^{-3}$ yr$^{-1}$ for initially coplanar systems, and $2.5$ Gpc$^{-3}$ yr$^{-1}$ for systems with randomly distributed inclinations. \citet{rodriguez_triple_2018} report a decrease in the estimated merger rates by a factor of $10$ by excluding triples with ZKL timescales smaller than $3$ Myr, postulating that these would have interacted and/or merged already before the formation of the BHs. This agrees with our simulations, which show that the vast majority of the inner stellar binaries experience mass transfer before the BHs form, unless the initial ZKL timescale exceeds that the nuclear timescales of the massive stellar progenitors. 

\begin{table}[]
\caption{Estimated merger rates for the stable and dynamically unstable systems.}
\label{tab:merger_rates}
\begin{tabular}{lrll}
\hline
Population & \multicolumn{1}{l}{\# simulated}                      & $\epsilon_\text{formation}$                                               & \begin{tabular}[c]{@{}l@{}}Merger rate\\ {[}Gpc$^{-3}$ yr$^{-1}${]}\end{tabular}                                         \\ \hline
Stable     & \cellcolor[HTML]{FFFFFF}{\color[HTML]{000000} 52464}  & \cellcolor[HTML]{FFFFFF}{\color[HTML]{000000} $6\times 10^{-7}$}            & \cellcolor[HTML]{FFFFFF}{\color[HTML]{000000} 4.99} \\
Unstable   & \cellcolor[HTML]{FFFFFF}{\color[HTML]{000000} 524640} & \cellcolor[HTML]{FFFFFF}{\color[HTML]{000000} $1.6 \times 10^{-7}$} & \cellcolor[HTML]{FFFFFF}{\color[HTML]{000000} 1.42} \\ \hline
\end{tabular}
\tablefoot{The number of simulated unstable systems is inflated by a factor of 10, as described in Sect. (\ref{sec:method_syzygy}). The formation efficiency $\epsilon_\text{formation}$ gives the number of ZAMS systems that end up in the merger channel as a fraction of all formed ZAMS systems.}
\end{table}

In these simulations we did not include natal supernova velocity kicks. Earlier studies have shown various effects of the natal kick on the final rates of BHB mergers in triples. While \citep{silsbee_lidov-kozai_2017} report a decrease in estimated merger by a factor of $40$ even with modest natal kicks, both \citep{silsbee_lidov-kozai_2017} and \citet{rodriguez_triple_2018} demonstrate that kicks only reduce merger rates by a factor of $2-4$, similar to the results from \citet{antonini_binary_2017}). Uncertainties in the physics of supernova kicks, emphasized by the discrepancies in its estimated effect on BHB mergers in triples, is one reason why we excluded this effect from our simulations. 

Our method for spin evolution during the n-body simulation does not include any spin-orbit interaction. For the dynamically stable systems, the high angular momentum in the inner orbit during ZKL oscillations do not result in any significant changes that would alter the outcomes. However, since the unstable triples are more chaotic, even small perturbations could potentially modify the evolution of these systems. Whether or not this would have a noticeable effect on the merger rates or merger properties likely depends on the specific dynamics experienced by the unstable systems. The systems that remain quasi-hierarchical while undergoing extreme ZKL oscillation would potentially be less susceptible to back-reactions from the spins, while the unstable triples that disintegrate and undergo democratic interactions are more prone to smaller perturbations. The overall effect of spin-orbit interactions on unstable triples is beyond the scope of this study but is an important and interesting subject for future work. 

\subsection{Comparison with previous studies}

\citet{silsbee_lidov-kozai_2017} find that a few percent of their merging systems have eccentricities greater than $0.999$ when entering the $10$ Hz GW frequency band. A similar result is reported by both \citet{dorozsmai_hierarchical_2025} and \citet{antonini_binary_2017}, with the latter finding that around 3\% of all their merging BHBs have eccentricities $e_\text{10Hz} \geq 0.999999$. In this study we find no systems with extreme eccentricities ($e \geq 0.9$) in the 10 Hz band. We do find that tighter inner orbits are more prone to having a higher eccentricity at merger (Table \ref{tab:spearman}). Different assumptions about the modelling of stellar evolution (maximum radii) and supernovae can both contribute to different orbits at the time of BHB formation. Given how rare the extreme-eccentricity binaries appear in the aforementioned studies, it is also simply possible that the lower number of systems simulated in this work ($N=52464$) compared to \citet{dorozsmai_hierarchical_2025} ($N=10^6$) include any triples that would in theory end up in this category. If we assume that $2-3\%$ of all BHBs merging via the triple channel have $e \geq 0.9$, we would expect only four to six such systems in this study, likely falling within the range of inherent uncertainties.

Several studies have also examined the evolution of black hole spin vectors in triple-induced mergers, mainly using orbit-averaged methods \citep{liu_spin-orbit_2017, antonini_precessional_2018, liu_black_2018, rodriguez_triple_2018, yu_spin_2020}. These studies have shown that the black hole spins in the inner binary slowly precess towards an attractor point, eventually freezing out as the inner binary shrinks enough to decouple from the tertiary. The location of this attractor point depends on the initial spin-orbit angles. Consequently, the final effective spin can have a wide range of values at the point of merger. We find a similar result for the final effective spin, but for a different reason. Slow precession towards an attractor point is not observed in our simulations. Instead, we find a random final spin-orbit misalignment due to the ZKL-induced precession of the inner angular momentum. These discrepancies could potentially occur as a result of different orbital configurations; we are simply not examining the same parameter space. 

A likely reason for our lack of slowly precessing systems is that other studies considered triples with smaller initial $a_\text{in}$ and $a_\text{out}$. For example, \citet{liu_spin-orbit_2017} examined highly compact triples with $a_\text{in} = 0.1$au and $a_\text{out} = 3,5,10$au. These systems occupy a completely separate part of the semi-major axis parameter space, while the ratio of orbital separations are similar to the triples in our work. Consequently, our inner binaries must dissipate substantially more orbital energy to merge. This indicates that these two populations undergo markedly different evolutions. With a wide inner binary, the energy dissipation required leads to two potential evolutionary channels. If the inner binary gradually shrinks through cycles of high eccentricity and dissipation, then each time the binary shrinks, the time until the next peak eccentricity increases ($t_\text{ZKL} \propto a_\text{out}/a_\text{in}$ \citep{kinoshita_analytical_1999}). The second channel is what we observe in this study, namely a rapid shrinkage of the inner orbit following extremely high eccentricity. The wide systems examined in this study may only permit the second channel, as the first would result in merger times exceeding either the allotted CPU time or the physical Hubble time. This could therefore explain the absence of systems undergoing slow precession of the spins towards an attractor point.

In addition, as reported by \citet{liu_black_2018}, the spin-orbit misalignment evolves toward the 'attractor' in systems where the octupole effect is negligible. In contrast, the triples with a more influential octupole effect (larger $\varepsilon_\text{oct}$) experience more chaotic evolution of the BH spins, producing a wide range of effective spins. With the majority of the triples in this study having large $\varepsilon_\text{oct}$ (Fig. \ref{fig:properties_at_ZAMS_and_BHBonset}, lower panel), it is highly likely that we exclude systems that experience slow decay into the attractor point.

\citet{rodriguez_triple_2018} considered triples with inner and outer orbit properties similar to those in this study, while also accounting for pre-BH stellar evolution. Differences with our study lie in the modelling of the stellar interactions and three-body dynamics. Whereas \citet{rodriguez_triple_2018} included mass-transferring systems (excluded here), we included three-body dynamics during the pre-BH evolution. Additionally, \citet{rodriguez_triple_2018} used the secular equations of motion to model the dynamics following the formation of an inner BHB and therefore omitted the most extreme eccentricity excitations that occur when these break down. These substantially differ from the assumption made in this study, and the inclusion of mass transfer provides \citet{rodriguez_triple_2018} with a population of more compact inner binaries with small eccentricities that reside in parts of the parameter space omitted in this study. With a more compact inner binary, there is a wider range of outer semi-major axes that give the triple a small octupole value; thus, we would expect more systems to experience spin angle freezing.

Finally, \citep{ginat_dynamical_2026} examined the dynamical evolution of so-called quasi-hierarchical triples. These are stable (hierarchical) triples in which the outer orbit has an eccentricity high enough for its outer periapsis to comparable in magnitude to $a_\text{in}$. They find that under certain conditions, the inner binary can reach eccentricities on quadrupole timescales that deviate from what is expected by traditional ZKL evolution, even for systems where the octupole term is zero. The authors report that this mechanism could enhance the ability of triples to produce GW mergers. In this study, the dynamically stable triples that produce a merging inner binary have an average $r_{\text{p, out}}/a_\text{in}$ of $10$; thus, the contribution of this mechanism to our overall merger rate is likely not significant.

\section{Conclusions}

In this study we examined the formation of BHB mergers due to eccentricity excitations in wide, non-interacting triples. We followed the evolution of massive triples from ZAMS to merger using a combination of population synthesis for stellar evolution, orbit-averaged equations for triple dynamics, and direct n-body simulations. To the best of our knowledge, this is the first report of both the estimated merger rates of black holes in dynamically unstable triples -- systems that were born stable but lost their hierarchy due to internal evolution -- as well as the result of evolving black hole spins using non-orbit-averaged techniques. Our results can be summarized as follows:

\begin{itemize}
    \item We find that about $1\%$ of triples with black hole progenitors result in a dynamically stable triple with an inner BHB, with the remaining systems primarily experiencing RLOF or disintegration before the black holes form. Similarly, we find that around $0.5\%$ of the triples produce an inner BHB but destabilize following a second supernova. Less than $0.1\%$ of the triples end up merging when evolved using orbit-averaged techniques.
    \item Simulating stable triples from the onset of BHB formation using n-body with post-Newtonian terms, we find that $0.03$\% of massive triples produce a BHB merger due to ZKL oscillations. This corresponds to an estimated merger rate density of $5$ Gpc$^{-3}$ yr$^{-1}$. 
    
    \item Likewise, about $0.01\%$ of massive triples produce BHB mergers after becoming dynamically unstable, which gives an estimated merger rate of $1.4$ Gpc$^{-3}$ yr$^{-1}$.
    
    \item Mutual inclination plays an important role in the formation of mergers in dynamically stable systems, with all the merging triples exhibiting inclinations close to $90^\circ$. This dependence is weaker for unstable triples, which show a substantially broader distribution in their initial inclination. Consequently, even triples born in coplanar orbits may still become sufficiently dynamically active to achieve a merger if they approach the limit of stability.
    \item Most of the merging binaries -- from both stable and unstable triples -- enter the 10 Hz GW frequency band with eccentricities in the range $(10^{-4}, 10^{-2})$. A small number of systems from the dynamically unstable population, however, have $e_\text{10Hz} \geq 0.03$. 
    \item The initial angle between the BH spin vectors and the inner binary angular momentum at the time of BHB formation does not seem to influence the final distribution of spin-orbit angles (or, consequently, the effective spin). During the ZKL phase, the BH spin vectors experience negligible precession, while the inner angular momentum can experience changes of the order of $180^\circ$ due to perturbations from the tertiary. The result is a chaotic evolution in the effective spins before the inner binary begins to merge. Once the inner periapsis becomes small enough to remove energy via GWs, the effective spin freezes out and remains at a constant value until merger. On a population level, this manifests as a broad distribution of effective spins that peaks at $0$. 
\end{itemize}

These results further emphasize that triples could be a substantial contributor to BHB mergers in the Universe, with reported merger rates up to $35\%$ of the current global BHB merger rate estimate \citep{abac_gwtc-40_2025}. In addition, the inner binaries that merge due to perturbations from a tertiary companion exhibit unique properties that can be observed by future GW observatories. This, in turn, can help shed light on the evolutionary and dynamical history of the most massive stars. Our work also underscores the importance of accounting for pre-BH stellar evolution. Unless the initial triple properties are just right, the three-body dynamics required to merge an inner binary are likely to induce stellar interaction before BHB formation, or the system may become unbound due to supernovae. Both outcomes highly restrict the properties a ZAMS triple system must have to eventually merge through this channel. Consequently, care must be taken when generating a population of triple BHs to ensure that a given system could evolve to host an inner BHB.

With the increasing sensitivity of current GW observatories, combined with planned instruments that will probe a larger portion of frequency space \citep{punturo_einstein_2010, amaro-seoane_laser_2017}, the future is bright not only for GW observations as a whole but also, specifically, for eccentric GW sources. New technologies will allow us to more easily detect signatures of eccentricity in GW signals and obtain more accurate measurements of BH spin properties. These will teach us more about the formation history of a merging BHB and, by extension, the evolutionary history of the stellar progenitors.

\section*{Data availability}
The data and code necessary to reproduce the figures in this paper are publicly available on Zenodo: \href{https://zenodo.org/records/18430115}{10.5281/zenodo.18430115}

\clearpage
\begin{acknowledgements}
TB is supported by the European Union’s Horizon Europe research and innovation programme under the Marie Sklodowska–Curie grant agreement No 101153423.\\
All the figures in this paper were made using the \texttt{Makie} visualization package \citep{DanischKrumbiegel2021}.

\end{acknowledgements}

\bibliographystyle{aa}
\bibliography{bibliography} 
\begin{appendix}

\section{The 2.5PN acceleration terms} \label{appendix:PN_acceleration}

We used the following equations to model general relativistic effects in the n-body simulation. The equations are taken from \citet{blanchet_post-newtonian_2024}. For the acceleration of a body $i$ due to the potential from a body $j$, we use the following notation. Absolute positions, velocities, and masses are defined as $\textbf{r}_i$, $\textbf{v}_i$, and $m_i$, respectively. The relative positions and velocities are written as $\textbf{r} = \textbf{r}_i - \textbf{r}_j$ and $\textbf{v} = \textbf{v}_i - \textbf{v}_j$ with their magnitudes defined as $r \equiv |\textbf{r}|$ and $v \equiv |\textbf{v}|$. The dot product between two vectors $\textbf{x}$ and $\textbf{y}$ is written as $(xy)$. Finally, the unit normal position vector between the two bodies is defined and written as $\textbf{n} \equiv \textbf{r}/r$. The full 1 to 2.5 PN acceleration can be written as

\begin{equation}
    \textbf{a}_\text{PN} = \frac{1}{c^2} \textbf{a}_\text{1PN} + \frac{1}{c^4} \textbf{a}_\text{2PN} + \frac{1}{c^5} \textbf{a}_\text{2.5PN} + \mathcal{O}(c^6),
\end{equation}

where

\begin{equation}
\begin{aligned}
    \textbf{a}_\text{1PN} = &\biggl[ \frac{5G^2 m_1 m_2}{r^3} + \frac{4G^2m_2^2} {r^3} + \frac{Gm_2}{r^2} \biggl( \frac{3}{2}(nv_2)^2 - v_1^2 \\
    &- 4(v_1v_2) - 2v_2^2 \biggr) \biggr] \textbf{n} + \frac{Gm_2}{r^2} (4(nv_1) - 3(nv_2))\textbf{v},
\end{aligned}
\end{equation}

\begin{equation}
\begin{aligned}
    \textbf{a}_\text{2PN} = &\biggl[ -\frac{57G^3 m_1^2 m_2}{4r^4} - \frac{69G^3m_2m_2^2}{2r^4} - \frac{9G^3 m_2^3}{r^4} \\ 
    &+ \frac{Gm_2}{r^2}\biggl(-\frac{15}{8}(nv_2)^4 
    + \frac{3}{2}(nv_2)^2v_1^2 - 6(nv_2)^2(v_1v_2) \\
    &- 2(v_1v_2)^2 + \frac{9}{2}(nv_2)^2v_2^2 + 4(v_1 v_2)v_2^2 - 2v_2^4 \biggr) \\
    &+ \frac{G^2m_1m_2}{r^3}\biggl(\frac{39}{2}(nv_1)^2 - 39(nv_1)(nv_2) \\
    &+ \frac{17}{2}(nv_2)^2 - \frac{15}{4}v_1^2 - \frac{5}{2}(v_1v_2) + \frac{5}{4}v_2^2 \biggr) \\
    &+ \frac{G^2 m_2^2}{r^3}(2(nv_1)^2 - 4(nv_1)(nv_2) - 6(nv_2)^2 \\
    &- 8(v_1v_2) + 4v^2) \biggr] \textbf{n} + \biggl[ \frac{G^2 m_2^2}{r^3}(-2(nv_1) - 2(nv_2)) \\
    &+ \frac{G^2 m_1 m_2}{r^3}\biggl( -\frac{63}{4}(nv_1) + \frac{55}{4}(nv_2) \biggr) \\
    &+ \frac{Gm^2}{r^2} \biggl(-6(nv_1)(nv_2)^2 + \frac{9}{2}(nv_2)^3 + (nv_2)v_1^2 \\
    &- 4(nv_1)(v_1v_2) + 4(nv_2)(v_1v_2) \\
    &+ 4(nv_1)v_2^2 - 5(nv_2)v_2^2 \biggr) \biggr] \textbf{v}
\end{aligned}
\end{equation}

\begin{equation}
\begin{aligned}
    \textbf{a}_\text{2.5PN} = &\biggl[ \frac{208G^3 m_1 m_2^2}{15r^4}(nv) - \frac{24G^3m_1^2 m_2}{5r^4}(nv) \\
    &+\frac{12G^2m_1m_2}{5r^3}(nv)v^2 \biggr] \textbf{n} + \biggl[\frac{8G^3 m_1^2 m_2}{5r^4} \\
    &- \frac{32G^3 m_1 m_2^2}{5r^4} - \frac{4G^2 m_1 m_2}{5r^3}v^2 \biggr]\textbf{v}.
\end{aligned}
\end{equation}

In a triple system with bodies $1$, $2$, and $3$, the total acceleration on body $1$ is then

\begin{equation*}
    \textbf{a}_1 = (\textbf{a}_\text{N}^{1,2} + \textbf{a}_\text{N}^{1,3}) + (\textbf{a}_\text{PN}^{1,2} + \textbf{a}_\text{PN}^{1,3}),
\end{equation*}

where $\textbf{a}_\text{N}^{1,i}$ is the Newtonian acceleration on body $1$ by body $i$:

\begin{equation}
\textbf{a}_\text{N}^{1,i} = -\frac{Gm_i}{|\textbf{r}_{1,i}|^3} \textbf{r}_{1,i}.
\end{equation}

\section{The spin precession equations} \label{appendix:spin_precession}

The following section gives the equations for the spin precession used to evolve the spins of the black holes. We utilize the same formalism as the previous section, with the addition of the spin of a body $i$ denoted as $\textbf{S}_1$, where $\textbf{S} \equiv c \textbf{S}_\text{true} = Gm^2 \bar{\chi}$. Here, $\bar{\chi}$ is the dimensionless spin parameter vector, whose magnitude $0 \leq |\bar{\chi}| \leq 1$. The total spin precession for object $1$ in the general frame is given as \citep{faye_higher-order_2006}

\begin{equation}
    \frac{d\textbf{S}_1}{dt} = \frac{1}{c^2} \textbf{T}_\text{1PN} + \frac{1}{c^3} \textbf{T}_\text{1.5PN} + \frac{1}{c^4} \textbf{T}_\text{2PN} + \mathcal{O}(c^{-5}),
\end{equation}

where the $1$ and $2PN$ terms are spin-orbit terms and are given by

\begin{equation}
\begin{aligned}
    \textbf{T}_\text{1PN} = \frac{Gm_2}{r^2}\left[ \textbf{S}(nv) - 2\textbf{n}(vS_1) + (\textbf{v}_1 - 2\textbf{v}_2(nS_1) \right],
\end{aligned}
\end{equation}

\begin{equation}
\begin{aligned}
    \textbf{T}_\text{2PN} = &\biggl\{ \textbf{S}_1 \biggl[(nv_2)(vv_2) - \frac{3}{2}(nv_2)^2(nv) \\
    &+ \frac{Gm_1}{r}\Bigl(-16(nS_1)(nv) + 3(v_1S_1) - 7(v_2S_1) \Bigr) \\
    &+ 2(nS_1)\frac{Gm_2}{r}(nv) \biggr] - \textbf{v}_1 \biggl[ \frac{3}{2}(nS_1)(nv_2)^2 \\
    &+ (vS_1)(nv_2) - (nS_1)\frac{G}{r}(6m_1 - m_2) \biggr] \\
    &+ \textbf{v}_2 \biggl[(nS_1)\Bigl(2(vv_2) + 3(nv_2)^2 \Bigr) \\
    &+ 2(nv)\Bigl((v_1S_1) + (v_2S_1) \Bigr) - 5(nS_1)\frac{G}{r}(m_1 - m_2)\biggr] \biggr\}
\end{aligned}.
\end{equation}

The $1.5$PN spin-spin term is given by, for example, \citep{buonanno_detecting_2003}

\begin{equation}
    \textbf{T}_\text{1.5PN} = -\frac{G}{r^3}(\textbf{S}_2 - 3(nS_2)\textbf{n} \times \textbf{S}_1).
\end{equation}

\section{Evolution of example systems}

\begin{table*}[]
\caption{Properties of example triple systems visualized in Figs. (\ref{fig:example_separations}) and (\ref{fig:example_L_angles})}
\label{tab:example_proprties}
\centering    
\begin{tabular}{@{}lllllllll@{}}
\toprule
ID & Unstable & $\log_{10}{a_\text{in}}$ [R$_\odot$] & $\log_{10}{a_\text{out}}$ [R$_\odot$] & $i_\text{mut}$ [deg] & $q_\text{in}$ & $q_\text{out}$ & $\log_{10}{(1 - e_\text{max})}$ & $e_\text{10Hz}$ \\ \midrule
1  & True     & 4.37                     & 5.47                      & 90.0           & 1.78          & 0.54           & -7.2                          & 0.85                      \\
2  & True     & 5.89                     & 7.26                      & 82.96          & 1.39          & 1.02           & -6.45                         & 0.022                       \\
3  & True     & 4.94                     & 6.58                      & 170.54         & 1.31          & 0.21           & -6.93                         & 0.68                      \\
4  & False    & 5.9                      & 6.81                      & 86.42          & 0.58          & 0.49           & -6.65                         & 0.028                      \\
5  & True     & 5.65                     & 6.77                      & 90.0           & 1.94          & 0.08           & -6.61                         & 0.037                      \\
6  & True     & 5.28                     & 6.12                      & 115.88         & 2.34          & 0.39           & -6.84                         & 0.23                      \\ \bottomrule
\end{tabular}
\tablefoot{The systems are numbered column-wise, such that ID 1 refers to the top-left sub-figure, ID 3 to the bottom-left, and ID 4 to the top-right.}
\end{table*}

\begin{figure*}
\centering
    \includegraphics[width=0.8\textwidth]{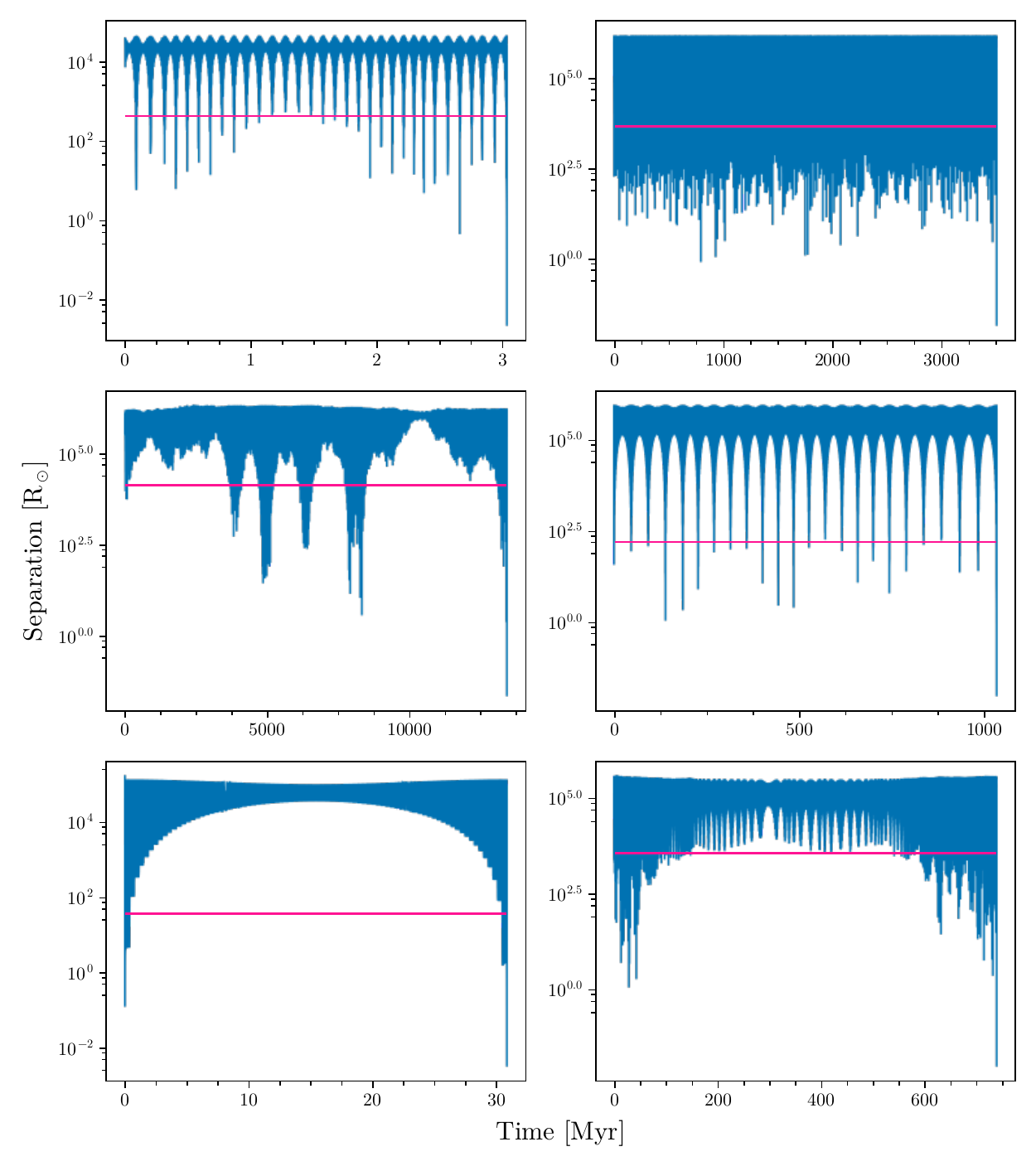}
    \caption{Examples of separations between the two inner components, from BHB formation to merger. The pink line shows the separation that determines the semi-secular regime, as defined by Eq. (\ref{eq:semisec}).}
    \label{fig:example_separations}
\end{figure*}

\begin{figure*}
\centering
    \includegraphics[width=0.8\textwidth]{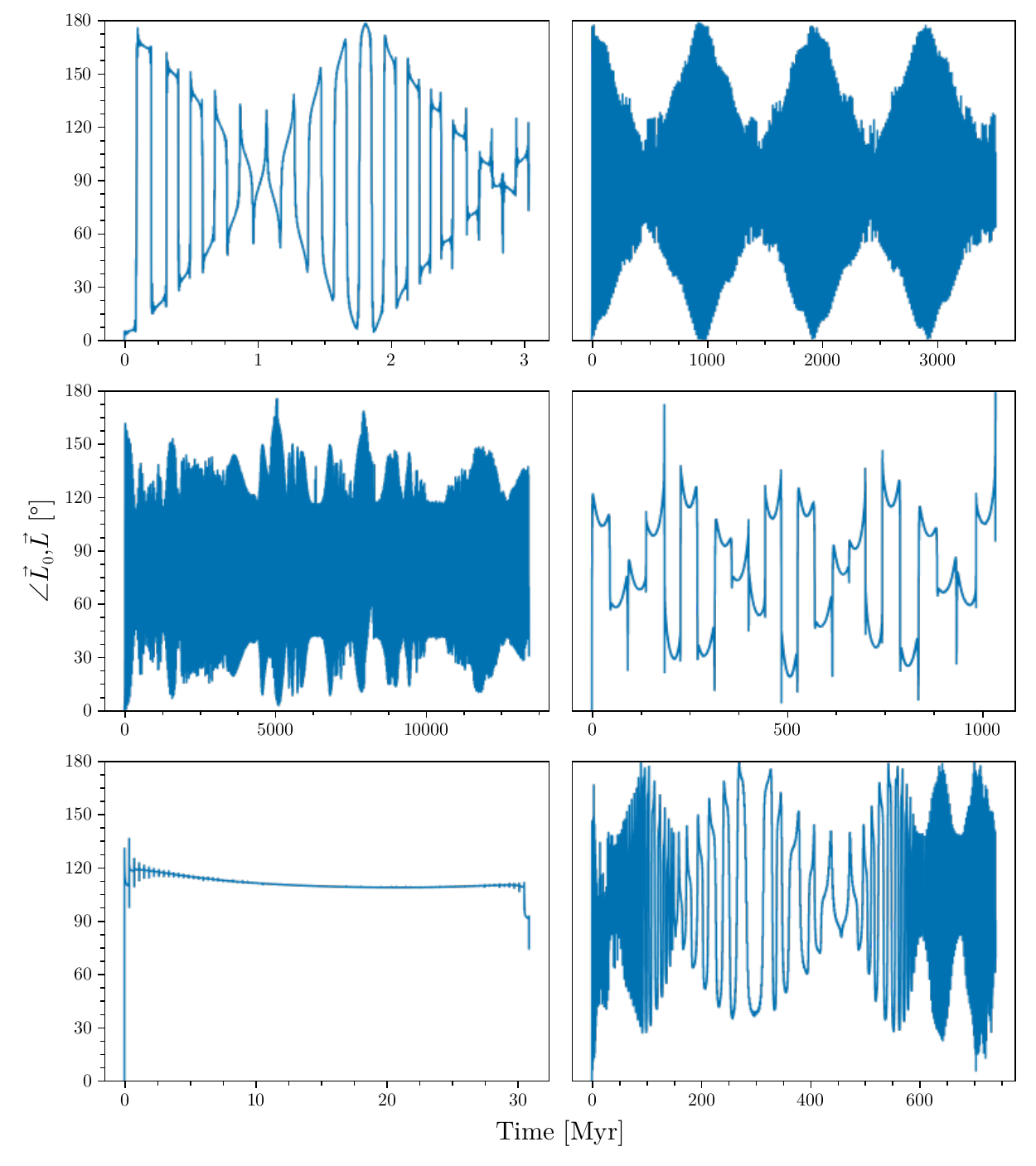}
    \caption{Angle between the angular momentum $\vec{L}(t)$ and $\vec{L}(t=0)$ of six different triples where the inner binary results in a merger.}
    \label{fig:example_L_angles}
\end{figure*}

In Figs. (\ref{fig:example_separations}) and (\ref{fig:example_L_angles}) we show the detailed evolution of six example systems that end up in a merger. The first figure shows the separation between the bodies in the inner binary as a function of time, while the second shows the angle between the inner angular momentum vector at a time $t$ and at time of BHB formation. The initial properties of the example systems are shown in Table (\ref{tab:example_proprties}).

\end{appendix}
\end{document}